\documentclass[11pt,a4paper]{article}
\usepackage{aas_macros}
\usepackage{jcappub}
\usepackage{amsmath}
\usepackage{amssymb}
\usepackage{graphicx}
\usepackage{epstopdf}
\usepackage{verbatim}
\usepackage{rotating}
\usepackage{multirow}
\usepackage[letterpaper]{geometry}

\newcommand{\vesc}{v_\textrm{esc}}
\newcommand{\vLSR}{v_\textrm{LSR}}
\newcommand{\vDg}{v_\textrm{D}}
\newcommand{\vlag}{v_\textrm{lag}}
\newcommand{\vSg}{v_\textrm{S}}
\newcommand{\vC}{v_\textrm{C}}
\newcommand{\lLSR}{l_\textrm{LSR}}
\newcommand{\bLSR}{b_\textrm{LSR}}
\newcommand{\llab}{l_\textrm{lab}}
\newcommand{\blab}{b_\textrm{lab}}
\newcommand{\lD}{l_\textrm{D}}
\newcommand{\bD}{b_\textrm{D}}
\newcommand{\lS}{l_\textrm{S}}
\newcommand{\bS}{b_\textrm{S}}
\newcommand{\lC}{l_\textrm{C}}
\newcommand{\bC}{b_\textrm{C}}
\newcommand{\AC}{A_\textrm{C}}
\newcommand{\AS}{A_\textrm{S}}
\newcommand{\AD}{A_\textrm{D}}
\newcommand{\Nesc}{N_\textrm{esc}}
\newcommand{\vg}{v_\textrm{g}}
\newcommand{\vq}{v_\textrm{q}}
\newcommand{\vqmin}{v_\textrm{q,min}}
\newcommand{\vqmax}{v_\textrm{q,max}}
\newcommand{\vlab}{v_\textrm{lab}}
\newcommand{\fg}{f_\textrm{g}}
\newcommand{\fgTM}{f_\textrm{g}^{\textrm{TM}}}
\newcommand{\mchi}{m_\chi}

\newcommand{\mN}{m_\textrm{N}}

\newcommand{\sigmavH}{\sigma_\textrm{H}}
\newcommand{\sigmavD}{\sigma_\textrm{D}}
\newcommand{\sigmavS}{\sigma_\textrm{S}}
\newcommand{\sigmavC}{\sigma_\textrm{C}}
\newcommand{\thetav}{\theta_\textrm{v}}
\newcommand{\phiv}{\phi_\textrm{v}}
\newcommand{\thetaCM}{\theta_\textrm{CM}}
\newcommand{\Npix}{N_\textrm{pix}}
\newcommand{\Nbins}{N_\textrm{bins}}
\newcommand{\Nsig}{N_\textrm{sig}}
\newcommand{\Nbg}{N_\textrm{bg}}
\newcommand{\Ntot}{N_\textrm{tot}}
\newcommand{\Emin}{E_\textrm{min}}
\newcommand{\Emax}{E_\textrm{max}}
\newcommand{\Rbg}{R_\textrm{bg}}
\newcommand{\dO}{d\Omega_\textrm{q}}

\title{Probing the Local Velocity Distribution of WIMP Dark Matter with Directional Detectors}
\author[a,b]{Samuel K.\ Lee}
\author[c]{Annika H. G. Peter}
\affiliation[a]{California Institute of Technology,\\Mail Code 350-17, Pasadena, CA
91125, USA}
\affiliation[b]{Department of Physics and Astronomy, Johns Hopkins University,\\Baltimore,
MD 21218, USA}
\affiliation[c]{Department of Physics and Astronomy, University of California,\\ Irvine, CA  92697-4575, USA}

\emailAdd{sklee@astro.caltech.edu}
\emailAdd{annika.peter@uci.edu}
\date{\today}

\abstract{We explore the ability of directional nuclear-recoil detectors to constrain the local velocity distribution of weakly interacting massive particle (WIMP) dark matter by performing Bayesian parameter estimation on simulated recoil-event data sets.  We discuss in detail how directional information, when combined with measurements of the recoil-energy spectrum, helps break degeneracies in the velocity-distribution parameters.  We also consider the possibility that velocity structures such as cold tidal streams or a dark disk may also be present in addition to the Galactic halo.  Assuming a $\mathrm{CF}_4$ detector with a 30-kg-yr exposure, a 50-GeV WIMP mass, and a WIMP-nucleon spin-dependent cross-section of $10^{-3}~\mathrm{pb}$, we show that the properties of a cold tidal stream may be well constrained.  However, measurement of the parameters of a dark-disk component with a low lag speed of $\sim\!\!50~\mathrm{km/s}$ may be challenging unless energy thresholds are improved.}

\begin{document}
\maketitle
\flushbottom

\section{Introduction} \label{sec:intro}

Solid-state and liquid WIMP-dark-matter detectors designed to measure the energy of nuclear recoils from WIMP collisions are entering maturity on both theoretical and experimental fronts.  Indeed, a large number of theoretical studies have investigated the statistical power of these experiments to characterize WIMP dark matter \cite{green2010,shan2011a,akrami2011b,pato2011a,peter2011,arina2011,frandsen2012,strege2012}.  Furthermore, a variety of experiments \cite{giuliani2010,cdms2010,behnke2011,felizardo2011,sekiya2011,ardm2011,ahmed2011,aprile2011,aprile2011b,malling2011,baudis2012} are currently running, with a few tantalizing signals already observed \cite{bernabei2010,aalseth2011,angloher2011}.  In contrast, gas detectors with sensitivity to the nuclear-recoil \emph{direction} via the measurement of ionization tracks are still relatively nascent.  Nevertheless, some theoretical studies on directional dark-matter detection have likewise been conducted \cite{copi1999,copi2001,morgan2005a,morgan2005b,copi2007,green2008,alenazi2008, finkbeiner2009,ahlen2010,green2010a,lisanti2010,billard2010,billard2010a,billard2011,billard2012,chiang2012}, and a small number of directional detectors are currently under development \cite{grignon2009,pipe2010,miuchi2010,ahlen2011,naka2011,vahsen2012}.

A primary advantage of these directional detectors is that they allow the possibility of easily distinguishing between terrestial background events (which should be isotropic) and WIMP-induced recoil events (which should be non-isotropic, due to our motion through the Galactic halo).  Perhaps even more intriguing is the possibility that directional detectors may allow the details of the local WIMP velocity distribution to be inferred.  Theoretical expectations and N-body simulations of the Galactic halo both give us reason to believe that even if the local spatial distribution of dark matter might be expected to be relatively smooth, the \emph{velocity} distribution may possess interesting structure.  Possibilities include cold tidal streams passing through the local solar neighborhood, a dark-matter disk aligned with the stellar disk, and warm debris flows \cite{lake1989,gelmini2004,freese2005,savage2006,read2008,vogelsberger2009,read2009,kuhlen2010,lang2010,
alves2010,lisanti2011a,maciejewski2011,natarajan2011,fantin2011,kuhlen2012,vergados2012}.  Directional-detection experiments may confirm this picture, perhaps shedding light on not only the local velocity distribution but also the process of structure formation on galactic scales.  Furthermore, a better understanding of the dark-matter velocity distribution will yield improved constraints on the particle properties of the dark matter -- in particular,  the WIMP mass and the WIMP-nucleon cross section.

In this paper, we explore the statistical ability of directional detectors to constrain the local WIMP velocity distribution.  The organization of the paper is as follows.  In section~\ref{sec:formalism}, we briefly review the general formalism for obtaining the directional recoil spectrum from the dark-matter velocity distribution.  We then discuss how a binned likelihood analysis of the directional recoil-event data may be used to estimate the parameters of the velocity distribution.  In section~\ref{sec:analyses}, we perform parameter estimation on simulated data sets to demonstrate the power of these methods.  We consider three specific distributions as examples: (1) the standard halo model, (2) a halo model with an additional cold dark-matter stream component, and (3) a halo model with an additional dark-matter disk component.  We discuss implications for future directional dark-matter-detection studies and give our conclusions in section~\ref{sec:conclusions}.

\section{Formalism} \label{sec:formalism}

\subsection{The Directional Recoil Spectrum}

The directional dependence of the WIMP signal was first discussed in ref.~\cite{spergel1988}.  The formalism presented here for the calculation of the nuclear-recoil event rate (as a function of recoil energy and direction) induced by WIMPs from a given WIMP velocity distribution was worked out in ref.~\cite{gondolo2002}.

Let the distribution of WIMP velocities $\bold{\vg}$ in the Galactic rest frame be given by $\fg(\bold{\vg})$.  If we neglect the gravitational influence of solar-system bodies, the distribution of WIMP velocities $\bold{v}$ in the lab frame moving with velocity $\bold{\vlab}$ with respect to the Galactic frame is then
\begin{equation}
f(\bold{v}) = \fg(\bold{v} + \bold{\vlab})\,,
\end{equation}
since the various velocities are related by $\bold{v} = \bold{\vg} - \bold{\vlab}$.

It can be shown that the directional recoil rate is given by
\begin{equation} \label{eq:spectrum}
\frac{dR}{dE \dO} = \frac{\rho_0 \sigma_N S(q)}{4\pi \mchi \mu_N^2} \widehat{f}(\vq, \widehat{\bold{q}})\,.
\end{equation}
Here, $R$ is the number of events per exposure (detector mass multiplied by time), $\rho_0$ is the local WIMP density, $\mu_N = \mchi \mN / (\mchi + \mN)$ is the reduced mass of the WIMP-nucleus system, $q\widehat{\bold{q}}$ is the lab-frame nuclear-recoil momentum, $E = q^2/2\mN$ is the lab-frame nuclear-recoil energy, $\vq = q/2\mu$ is the minimum lab-frame WIMP speed required to yield a recoil energy $E$, the WIMP-nucleus elastic cross section is $d\sigma/dq^2 = \sigma_N S(q) / 4 \mu_N^2 v^2$ (where $S(q)$ is the nuclear form factor), and 
\begin{equation} \label{eq:radon}
 \widehat{f}(\vq, \widehat{\bold{q}}) = \int\! \delta(\bold{v}\cdot\widehat{\bold{q}} - \vq) f(\bold{v})\, d^3v
\end{equation}
is the Radon transform of the lab-frame WIMP velocity distribution.  In practice, it is often easier to take the Radon transform of the Galactic-frame velocity distribution, and then use the relation
\begin{equation} \label{eq:radonprop}
\widehat{f}(\vq,\widehat{\bold{q}}) = \widehat{\fg}(\vq + \bold{\vlab}\cdot\widehat{\bold{q}},\widehat{\bold{q}})
\end{equation}
in  eq.~(\ref{eq:spectrum}) to find the directional recoil spectrum.  In this paper, we shall assume that the velocity of the lab frame is given by that of the local standard of rest (LSR); i.e., we take $\bold{\vlab} = (\vLSR, \lLSR, \bLSR) = (220~\mathrm{km/s}, 90^\circ, 0^\circ)$ in Galactic coordinates \cite{kerr1986}.  We shall ignore both the motion of the Earth around the Sun and the rotation of the Earth.

\subsection{The Binned Likelihood Function}

Our ultimate goal will be to investigate the degree to which a binned likelihood analysis of observed recoil events might recover the parameters of the dark-matter velocity distribution.  Given a number of observed events, we may construct a sky map of the data by binning both signal and background events into $\Npix$ pixels (we shall use the pixels of equal angular area given by \textsc{HEALPix} \cite{gorski2005} in this paper), as well as binning events in each pixel into $\Nbins$ energy bins (we shall also assume the energy bins are of equal width).  That is, the sky map specifies the number of observed signal and background events $M_{ij}$ in the $i$-th pixel of solid angle $d\Omega_i$ and the $j$-th energy bin of width $\Delta E_j$ for all $\Npix$ pixels.  We use pixels and bins as a proxy for finite angular and energy resolution in the detectors.

We would then like to construct a likelihood function that may be used to compare the observed sky map to the predicted sky map.  The predicted sky map can be specified by the total number of events $\Ntot$, which are observed over the energy-sensitivity range $[\Emin, \Emax]$, as well as the normalized distribution $P(\widehat{\bold{q}},E)$ in angle and energy of these events.  These quantities are simply related to the direction recoil spectrum via
\begin{equation} \label{eq:signalrelation}
\lambda \Ntot P(\widehat{\bold{q}},E) = \mathcal{E} \frac{dR}{dE \dO}\,,\,\, E \in [\Emin,\Emax]\,.
\end{equation}
Here, $\lambda$ is related to the background-rejection power of the experiment, defined so that the number of signal events is $\Nsig = \lambda\Ntot$ and the number of background events is $\Nbg = (1-\lambda)\Ntot$.   The effective exposure $\mathcal{E}$ gives the fraction of the total exposure $\mathcal{E}_\mathrm{tot}$ arising from the total mass of \emph{target} nuclei; we have implicitly assumed the detector acceptance is not energy dependent.  

Note that the predicted angular distribution of signal events is then given by a normalized integral over the energy-sensitivity range
\begin{eqnarray} \label{eqn:Pqhat}
P(\widehat{\bold{q}}) &=& \frac{\int_{\Emin}^{\Emax}\! dE \frac{dR}{dE \dO}}{\int\! \dO \int_{\Emin}^{\Emax}\! dE \frac{dR}{dE \dO}}\nonumber\\
		            &=& \frac{\int_{\vqmin}^{\vqmax}\! d\vq \vq S(2\mu_N\vq) \widehat{f}(\vq,\widehat{\bold{q}})}{\int\! \dO \int_{\vqmin}^{\vqmax}\! d\vq \vq S(2\mu_N\vq) \widehat{f}(\vq,\widehat{\bold{q}})}\,,
\end{eqnarray}
where $\Emin = 2\vqmin^2\mu_N^2/\mN$ and $\Emax = 2\vqmax^2\mu_N^2/\mN$.  The energy distribution $P(E)$ of signal events is similarly given by a normalized integral over all angles.

 For a given velocity distribution, we see that the predicted mean number of signal and isotropic background events in the $i$-th pixel (centered at the direction $\widehat{\bold{q}}_i$) and $j$-th energy bin is then given by
\begin{eqnarray}
\bar{M}_{ij} &=& \Ntot  \int_{d\Omega_i}\!\! d\Omega \int_{\Delta E_j}\!\! dE\, [\lambda P(\widehat{\bold{q}},E) + (1-\lambda)  \frac{P_B(E)}{4\pi}] \nonumber\\
	&\approx& \Ntot  d\Omega_i \int_{\Delta E_j}\! dE\, [\lambda P(\widehat{\bold{q}}_i,E) + (1-\lambda) \frac{ P_B(E)}{4\pi}]\,,
\end{eqnarray}
where $P_B(E)$ is the energy distribution of the isotropic background events, normalized to unity over the recoil-energy sensitivity range.  The approximation in the second line (which approximates the angular integral over each pixel with the value of the integrand at the center of each pixel multiplied by the pixel size) holds in the limit that $\Npix$ is large.  In each energy-binned pixel, the number of events is Poisson distributed, so a suitable likelihood function is given by the product of the distributions in each energy-binned pixel
\begin{equation}
\mathcal{L} = \prod_{i=1}^{\Npix}\prod_{j=1}^{\Nbins} \wp(M_{ij} | \bar{M}_{ij})\,,
\end{equation}
where $\wp(M_{ij} | \bar{M}_{ij})$ is the Poisson distribution function for the random variable $M_{ij}$ (the observed number of events) with mean $\bar{M}_{ij}$ (the predicted number of events, which depends on the velocity distribution $\fg$ and the experimental conditions).

Let us now assume that the Galactic-frame velocity distribution $\fg$ depends on the model parameters $\theta_k$, the measurement of which is of interest.  Then the Radon transform $\widehat{f}(\vq,\widehat{\bold{q}})$ of the lab-frame velocity distribution also depends on $\theta_k$.  From eq.~(\ref{eq:radonprop}), we see it further depends on $\bold{\vlab}$, which may be treated as three additional model parameters $\{\vlab,\llab,\blab\}$ (in Galactic coordinates).  Thus, it is clear that $P(\widehat{\bold{q}},E)$ depends on $\theta_k$ and $\bold{\vlab}$.  Furthermore, examining the form of $P(\widehat{\bold{q}},E)$ in eq.~(\ref{eqn:Pqhat}), we see that it additionally depends on $\mu_N$ through the argument of the form factor and the limits of the integrals.  Therefore, it depends on the known target-nucleus mass $\mN$ and the unknown WIMP mass $\mchi$, the latter of which may be treated as yet another model parameter we would like to measure.  However, the direct dependence of $P(\widehat{\bold{q}},E)$ on $\mchi$ weakens at high values of $\mchi$, since the dependence of $\mu_N$ on $\mchi$ is also weak at such values.  Finally, the background-rejection power of the detector may not be well known, so we may also treat $\lambda$ as a model parameter.  

The likelihood function $\mathcal{L}$ can therefore be written in terms of the functions and parameters specifying both the velocity distribution and the experimental conditions as $\mathcal{L}[$$\mchi$,$\lambda$,$\vlab$,$\llab$,$\blab$,$\theta_k$,$S(q)$; $m_N$,$\Emin$,$\Emax$,$\Npix$,$\Nbins$,$\Ntot$,$P_B(E)]$.  By assuming the forms of $S(q)$ and $P_B(E)$, a likelihood analysis of observed data can then be done to estimate the parameters that are unknown.  In this work, we shall take $S(q) = 1$ and assume that $P_B(E) = 1/(\Emax - \Emin)$ is a flat spectrum for simplicity.

Note that in this treatment, we have folded all the dependence of the \emph{amplitude} of the signal $\Nsig = \lambda\Ntot$ into $\lambda$.  However, from eqs.~(\ref{eq:spectrum})~and~(\ref{eq:signalrelation}), it is clear that $\Nsig$ will itself depend not only on the exposure $\mathcal{E}_\mathrm{tot}$, but also on the additional, totally degenerate parameters $\rho_0$ and $\sigma_N$.  If one is more interested in the particle properties of the WIMP, the usual procedure is to fix $\rho_0 \approx 0.3~\textrm{GeV}/\textrm{cm}^3$ (a typical estimate of the local dark-matter density, which may be measured by various means \cite{kuijken1989b,bergstrom1998b,holmberg2000,catena2010,iocco2011}) and to treat $\sigma_N$ as the parameter of interest.  In fact, more often the focus is placed on the WIMP-\emph{nucleon} cross section $\sigma_{p,n}$; for a spin-dependent interaction, this is related to the WIMP-nucleus cross section via
\begin{equation}
\frac{\sigma_N}{\sigma_{p,n}} = \frac{4}{3} \frac{\mu_N^2}{\mu_{p,n}^2} \frac{J+1}{J} \frac{(a_p\langle S_p \rangle+a_n\langle S_p \rangle)^2}{a_{p,n}^2}\,,
\end{equation}
where $\mu_{p,n}$ is the WIMP-nucleon reduced mass, $J$ is the nuclear spin, $a_{p,n}$ are the effective nucleon coupling strengths, and $\langle S_{p,n} \rangle$ are the expectation values of the spin content of the nucleon group \cite{jungman1996,giuliani2004}.

However, it can be argued that the typical value of $\rho_0 \approx 0.3~\textrm{GeV}/\textrm{cm}^3$ usually assumed may not even be relevant for direction-detection experiments; it is simply a large-scale average of the dark-matter density at the Galactic radius of the Sun $r_0 \approx 8.5~\textrm{kpc}$, and does not account for the possible existence of substructure in the immediate neighborhood of the Earth \cite{kamionkowski2008,vogelsberger2009}.  The assumption of this typical value then strongly colors any conclusions drawn about the estimated value of $\sigma_{p,n}$.  Thus, perhaps a more assumption-independent approach would be to ask: if we have observed a given number of events $\Ntot$, how well can the parameters of an assumed velocity distribution be estimated using only the \emph{distribution} $P(\widehat{\bold{q}},E)$ of these events?  In this way, we will sidestep the issues introduced by the degeneracy of $\rho_0$ and $\sigma_N$, which are now subsumed into the single parameter $\lambda$.

\section{Likelihood Analyses of Simulated Data} \label{sec:analyses}

We shall now explicitly demonstrate the feasibility of estimating the parameters of the velocity distribution by performing likelihood analyses on simulated data sets.  We shall consider three parameterized velocity distributions: 1) the standard halo model, 2) a halo model with an additional cold stream component, and 3) a halo model with a disk component.  The method of analysis is as follows.  Assigning fiducial values for the velocity-distribution and experimental parameters, we first randomly generate a number of recoil events using the procedure described in appendix~\ref{sec:eventgen}.  We then bin the simulated events in angle (using \textsc{HEALPix}) and in energy to create a simulated sky map.  We then use this to calculate the likelihood function, employing \textsc{MultiNest} \cite{feroz2008,feroz2009} to sample the likelihood function within the model parameter space, assuming flat priors.  The \textsc{getdist} routine from the \textsc{CosmoMC} package \cite{lewis2002} is then used to calculate the 1D and 2D marginalized posterior probability distributions.  We also calculate the minimum credible intevals (MCIs) (as defined in \cite{hamann2007}) of the posterior probability distributions, and examine how well the fiducial parameters have been recovered.

The fiducial values chosen for the velocity-distribution parameters will be discussed below for each of the three cases.  However, let us first motivate the choice of the values for the experimental parameters.  We shall perform the analyses assuming that the simulated data were collected by a $\mathrm{CF}_4$ MIMAC-like experiment \cite{santos2007,grignon2009,mayet2009,santos2011}.  That is, we assume the target nucleus is $^{19}F$, and that the relevant WIMP-nucleus interaction is spin-dependent and can be modeled using $J=1/2$, and assuming a pure proton coupling with $a_p = 1$, $a_n = 0$, and $\langle S_p \rangle = 0.5$.  Furthermore, we take the energy-sensitivity range to be 5--50 keV.  This range corresponds to that quoted by the MIMAC collaboration; the 5-keV threshold arises from the ionization threshold (taking into account quenching), while the upper bound is chosen to limit contamination from background events that dominate the signal at higher energies.  We shall take a fiducial value of $\mchi = 50~\mathrm{GeV}$, so we see that the 5 keV recoil-energy threshold corresponds to a sensitivity to WIMP velocities down to $\sim\!\!150~\mathrm{km/s}$.

The experimental parameters also include the angular and energy resolution.  In all of the analyses below in which we use directional information, we shall take the number of pixels to be the same, setting $\Npix = 768$ (i.e., \textsc{HEALPix} order 8).  This roughly corresponds to the $\sim\!\!10^\circ$ angular resolution expected to be attained by a MIMAC-like experiment.  On the other hand, we shall investigate the effect of varying the energy resolution by considering different numbers of energy bins $\Nbins$ for each analysis, which will be given in detail below.  For now, we note that the bin widths we shall assume are relatively large and conservative, considering that the energy resolution of the micromegas detectors used in the MIMAC experiment is expected to be $\sim\!\!15\%$.  In any case, the exact values assumed for the angular and energy resolutions do not have a large effect on the quality of the parameter estimation, and it can be shown that unbinned likelihood analyses yield similar results.

Finally, there remains the question of the number of signal and background events we should examine for each analysis.  We will assume various values of $\Nsig$ and $\Nbg$ for each case, as will be discussed and motivated below.

Values for the experimental parameters used for each analysis are summarized in table~\ref{tab:expparams}.    Fiducial values for the velocity-distribution parameters and the flat prior ranges are summarized in table~\ref{tab:fidparams}; we shall proceed to discuss the choice of these fiducial parameters for each velocity distribution in detail.

\begin{table}[t]
\begin{tabular}{l|lllllllll}
\hline\vspace{-.1cm}
\multirow{2}{*}{model} & \multirow{2}{*}{$\Npix$} & \multirow{2}{*}{$\Nbins$} & $\Emin$ & $\Emax$ & \multirow{2}{*}{$\Nsig$} & \multirow{2}{*}{$\Nbg$} & \multirow{2}{*}{$\Ntot$} & $\mathcal{E}_\mathrm{tot}$ & $\Rbg$ \\
& & & (keV) & (keV) & & & & (kg-yr) & ([kg-yr]$^{-1}$)\\
\hline\vspace{-.1cm}
halo-only & 768 & 10 &\multirow{2}{*}{5} & \multirow{2}{*}{50} & \multirow{2}{*}{100} & \multirow{2}{*}{0} & \multirow{2}{*}{100} & \multirow{2}{*}{4.4} & \multirow{2}{*}{0}  \\
($\mchi$ fixed) & (or 1) & (or 1) & & & & & & & \\
\hline\vspace{-.1cm}
halo-only & 768 & 10 &\multirow{2}{*}{5} & \multirow{2}{*}{50} & \multirow{2}{*}{100} & \multirow{2}{*}{0} & \multirow{2}{*}{100} & \multirow{2}{*}{4.4} & \multirow{2}{*}{0}  \\
($\mchi$ flat prior) & (or 1) & (or 1) & & & & & & & \\
\hline\vspace{-.1cm}
halo-only & 768 & \multirow{2}{*}{10} &\multirow{2}{*}{5} & \multirow{2}{*}{50} & \multirow{2}{*}{100} & \multirow{2}{*}{100} & \multirow{2}{*}{200} & \multirow{2}{*}{4.4} & \multirow{2}{*}{23} \\
(6 parameters) & (or 1) & & & & & & & &  \\
\hline
halo+stream & 768 & 20 & 5 & 50 & 650 & 300 & 950 & 28.9 & 10.4 \\
\hline
halo+disk & 768 & 36 & 5 & 50 & 541 & 272 & 813 & 20 & 15 \\
\hline\vspace{-.1cm}
halo+disk & \multirow{2}{*}{768} & \multirow{2}{*}{40} &\multirow{2}{*}{0} & \multirow{2}{*}{50} & \multirow{2}{*}{900} & \multirow{2}{*}{300} & \multirow{2}{*}{1200} & \multirow{2}{*}{20} & \multirow{2}{*}{15} \\
(zero threshold) & & & & & & & & &\\
\hline
\end{tabular}
\caption{Experimental parameters used to simulate data for each analysis.  The quoted values of $\mathcal{E}_\mathrm{tot}$ and $\Rbg$ assume a $\mathrm{CF}_4$ detector and the typical values $\rho_0 \approx 0.3~\mathrm{GeV/cm^3}$ and $\sigma_{p,n} \approx 10^{-3}~\textrm{pb}$.}
\label{tab:expparams}
\end{table}

\subsection{Halo-only Model}

For simplicity, we shall first consider a Galactic dark-matter halo with a velocity distribution that may be locally approximated as an isotropic Maxwellian with velocity dispersion $\sigmavH$, truncated at the Galactic escape speed $\vesc$.  This truncated-Maxwellian distribution is generally referred to as the standard halo model, and is of the form
\begin{equation}
\fgTM(\bold{\vg}; \sigmavH,\vesc) = \frac{1}{\Nesc (2\pi\sigmavH^2)^{3/2}} \exp\left(-\frac{\vg^2}{2\sigmavH^2}\right)\,\theta(\vesc-\vg)\,,
\end{equation}
where
\begin{equation}
\Nesc(\sigmavH,\vesc) = \mathrm{erf}\left(\frac{\vesc}{\sqrt{2}\sigmavH}\right) - \sqrt{\frac{2}{\pi}}\frac{\vesc}{\sigmavH} \exp\left(-\frac{\vesc^2}{2\sigmavH^2}\right)\,.
\end{equation}
The Radon transform of this distribution is given by
\begin{eqnarray}
\widehat{\fgTM}(w,\bold{\widehat{w}}; \sigmavH,\vesc) &=& \int\! \delta(\bold{\vg}\cdot\widehat{\bold{w}} - w) \fgTM(\bold{\vg; \sigmavH,\vesc})\, d^3\vg\\ &=&\frac{1}{\Nesc (2\pi \sigmavH^2)^{1/2}} \left[\exp\left(-\frac{w^2}{2\sigmavH^2}\right)-\exp\left(-\frac{\vesc^2}{2\sigmavH^2}\right)\right]\theta(\vesc-w)\,.
\end{eqnarray}
The lab frame moves with respect to this velocity distribution with a velocity $\bold{\vlab}$, so as in eq.~(\ref{eq:radonprop}) we may find the Radon transform of the velocity distribution in the lab frame
\begin{equation}
\widehat{f^\textrm{H}}(\vq,\widehat{\bold{q}}) = \widehat{\fgTM}(\vq + \bold{\vlab}\cdot \bold{\widehat{q}}, \bold{\widehat{q}}; \sigmavH,\vesc)\,,
\end{equation}
which yields the directional recoil spectrum via eq.~(\ref{eq:spectrum}).

Note that integration of $\widehat{f^\textrm{H}}(\vq,\widehat{\bold{q}})$ over angles gives the recoil spectrum in the usual way; this integral takes the simple analytic form \cite{lewin1996}
\begin{equation} \label{eq:halospectrum}
\frac{dR}{dE} \propto \frac{\sqrt{\pi}}{2\sqrt{2}} \frac{\sigmavH}{\vlab}\left\{ \mathrm{erf}\left[\frac{\vq(E)+\vlab}{\sqrt{2}\sigmavH}\right] - \mathrm{erf}\left[\frac{\vq(E)-\vlab}{\sqrt{2}\sigmavH}\right]\right\} - \exp\left(-\frac{\vesc^2}{2\sigmavH^2}\right)\,,
\end{equation}
which approaches a falling exponential in the limit that $\vlab \to 0$ and $\vesc \to \infty$.

For the standard halo model, we see that the set of parameters determining the Galactic-frame velocity distribution is simply $\theta_k = \{\sigmavH,\vesc\}$.  Here and afterwards, we shall assume $\vesc = 550$ km/s is known independently (and in practice, the energy-sensitivity range is such that the analysis is not sensitive to the exact value of $\vesc$).  In total, there are then six parameters ${\{\mchi, \lambda, \vlab, \llab, \blab, \sigmavH\}}$ that will determine the energy and angular distribution of events.  In the following analyses, we shall consider a standard halo model with fiducial parameter values $\{\mchi = 50~\textrm{GeV}$, $\vlab = 220~\textrm{km/s}$, $\llab = 90^\circ$, $\blab = 0^\circ$, $\sigmavH = \vlab/\sqrt{2} = 155~\textrm{km/s}\}$.  Note that the choice of $\sigmavH = \vlab/\sqrt{2}$ yields a velocity distribution corresponding to a singular isothermal sphere, with halo profile $\rho(r) \approx \rho_0 (r_0/r)^{2}$.

\subsubsection{$\vlab-\sigmavH$ Analyses} \label{subsec:23param}

\begin{figure}[p]
\centering
$\begin{array}{cc}
\parbox[c]{3in}{\includegraphics[width=3in]{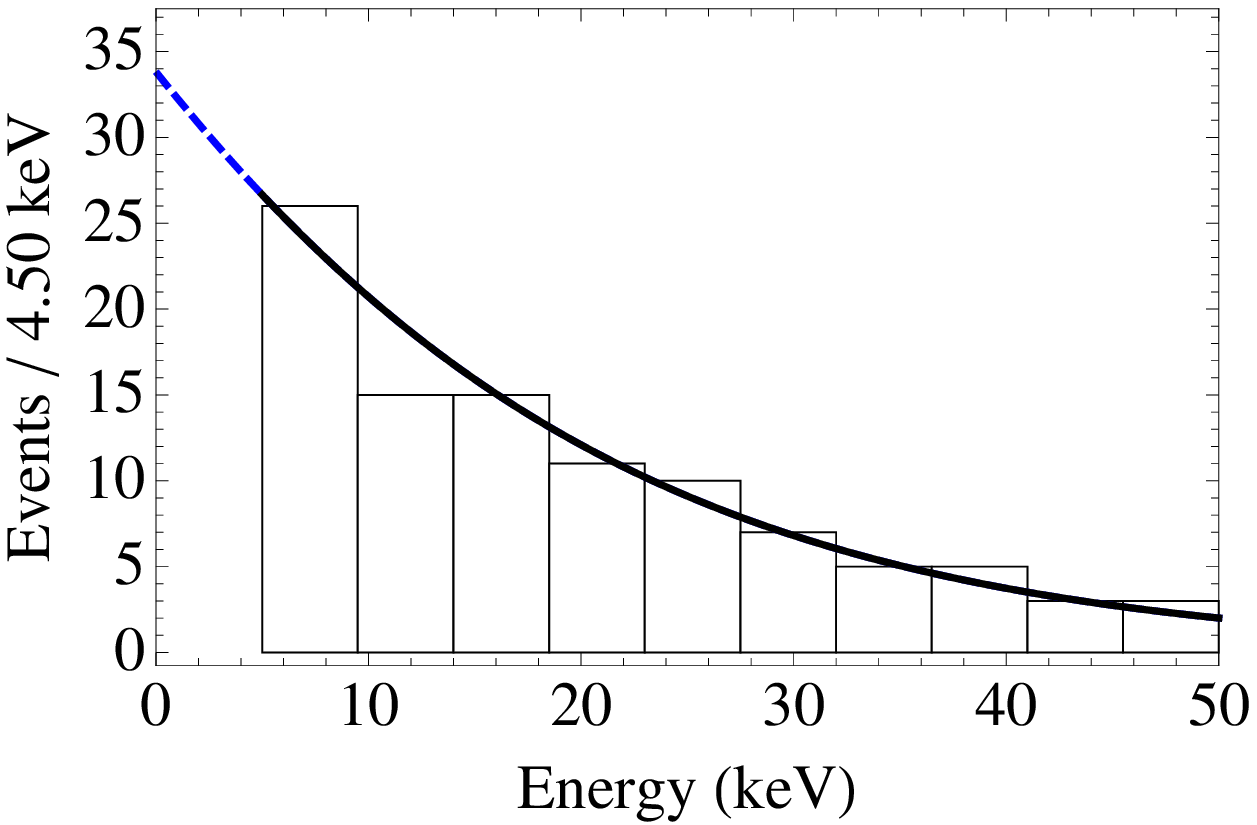}} &
\parbox[c]{3in}{\includegraphics[width=3in]{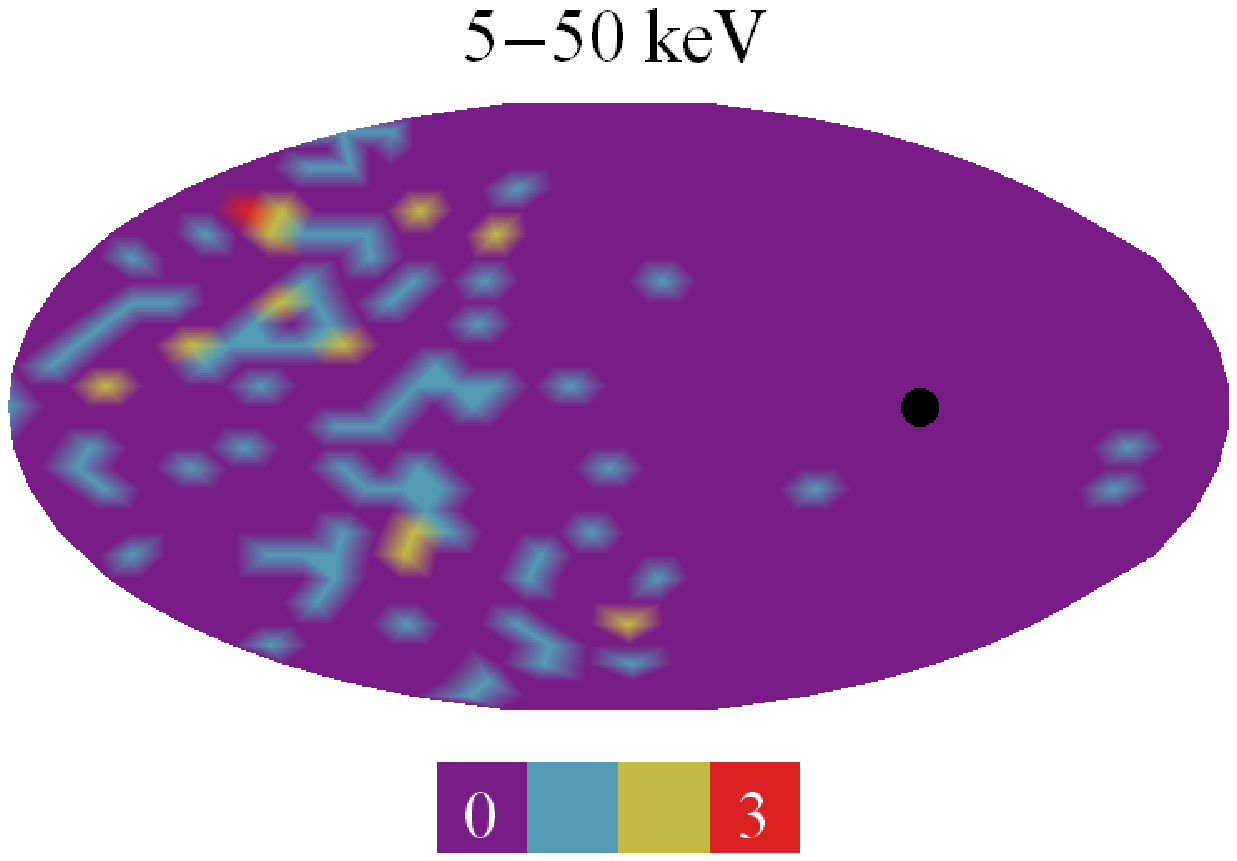}}
\end{array}$
\caption{Left: Simulated recoil spectrum for the halo-only 2-parameter and 3-parameter analyses in section~\ref{subsec:23param}, with binned signal events.  The halo-only spectrum calculated from the fiducial values is also plotted, and is shown in solid black inside the energy sensitivity range.  Right: Simulated recoil map for these analyses, in Mollweide projection.  The black dot indicates the direction of the LSR, given in Galactic coordinates by $(\lLSR, \bLSR) = (90^\circ, 0^\circ)$.}
\label{fig:vonly-map-sp}
\end{figure}

To first gain a qualitative and intuitive understanding of the additional power that directional information provides, let us perform a simple illustrative exercise.  Using the fiducial parameter values defining the standard halo model above, we generate $\Nsig = 100$ signal and $\Nbg = 0$ background events (i.e., we take $\Ntot = 100$ and $\lambda = 1$).  The spectrum and recoil map for these events is shown in figure~\ref{fig:vonly-map-sp}.

We then further assume that the values of the parameters ${\{\mchi, \lambda, \llab, \blab\}}$ are known exactly, so that only $\{\vlab,\sigmavH\}$ are unknown and remain to be estimated.  Sampling the likelihood function over the 2D $\{\vlab,\sigmavH\}$ parameter space using $\textsc{MultiNest}$ (assuming flat priors over the ranges given in table~\ref{tab:fidparams}), we then perform three separate likelihood analyses of these 100 events.  For the first analysis, we use only the energy information of the recoil events, accomplished by setting $\Npix = 1$ and $\Nbins = 10$.  For the second analysis, we use only the directional information, setting $\Npix = 768$ and $\Nbins = 1$.  Finally, we analyze the data using both direction and energy information, setting $\Npix = 768$ and $\Nbins = 10$.  Doing these three separate analyses allows us to study exactly how the direction and energy information translate into information about $\vlab$ and $\sigmavH$.

\begin{figure}[p]
\centering
\includegraphics{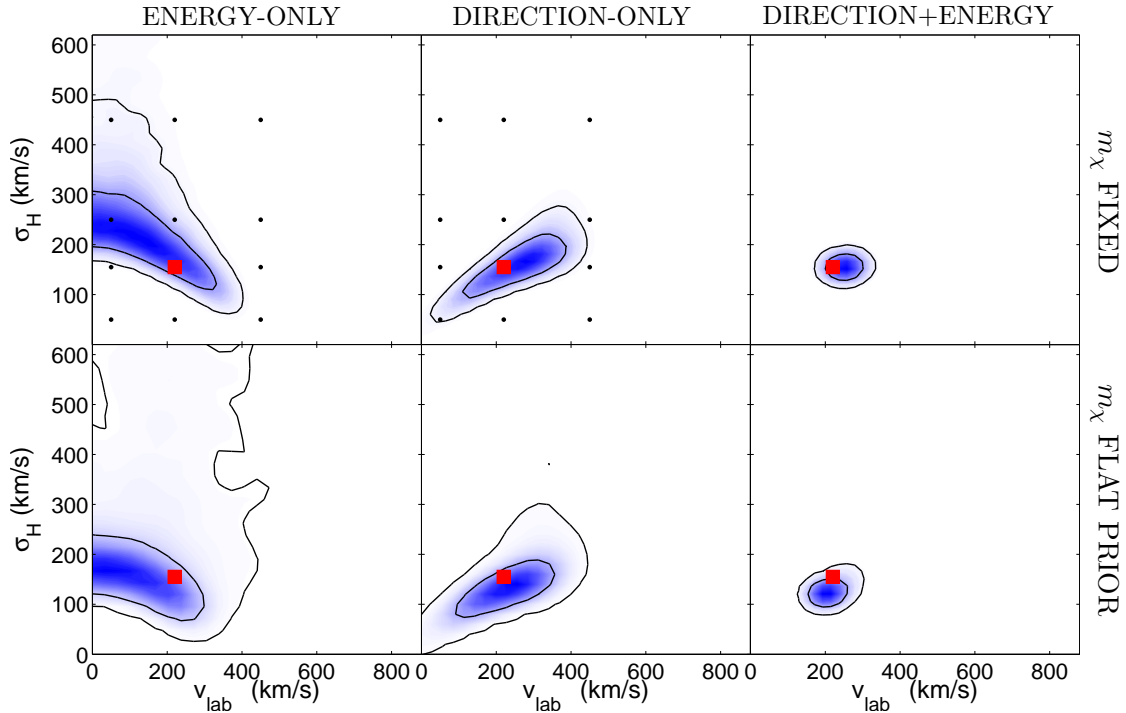}
\caption{Top row: Contour plots for the 2D posterior probability distribution in $\vlab$-$\sigmavH$ space, for the halo-only 2-parameter analyses with fixed $\mchi$ in section~\ref{subsec:23param}. Analyses using energy-only, direction-only, and direction+energy information are shown.  Red square markers indicate the fiducial values used in simulating the data.  Black dots indicate the values used to generate the spectra and maps in figures~\ref{fig:vonly-grid-sp}~and~\ref{fig:vonly-grid-map}.  The marginalized posterior probability is shaded blue, with contours indicating $68\%$ and $95\%$ confidence levels.  Bottom row: The same for the 3-parameter analyses assuming a flat mass prior.  Note that the energy-only contours are larger in the $\sigmavH$ direction when only a flat prior on $\mchi$ is known instead of the exact value; in contrast, the direction+energy contours are relatively insensitive to lack of knowledge about $\mchi$.}
\label{fig:vonly-tri}
\end{figure}

\begin{figure}[p]
\centering
$\begin{array}{c@{\hspace*{-23px}}c@{\hspace*{-21px}}cc}
\multicolumn{1}{c}{\normalsize{\hspace{3.5em}\vlab = 50~\mathrm{km/s}}} & \multicolumn{1}{c}{\normalsize{\hspace{2.75em}\vlab = 220~\mathrm{km/s}}} & \multicolumn{1}{c}{\normalsize{\hspace{.75em}\vlab = 450~\mathrm{km/s}}} & \\
\parbox[c]{130px}{\includegraphics[width=130px]{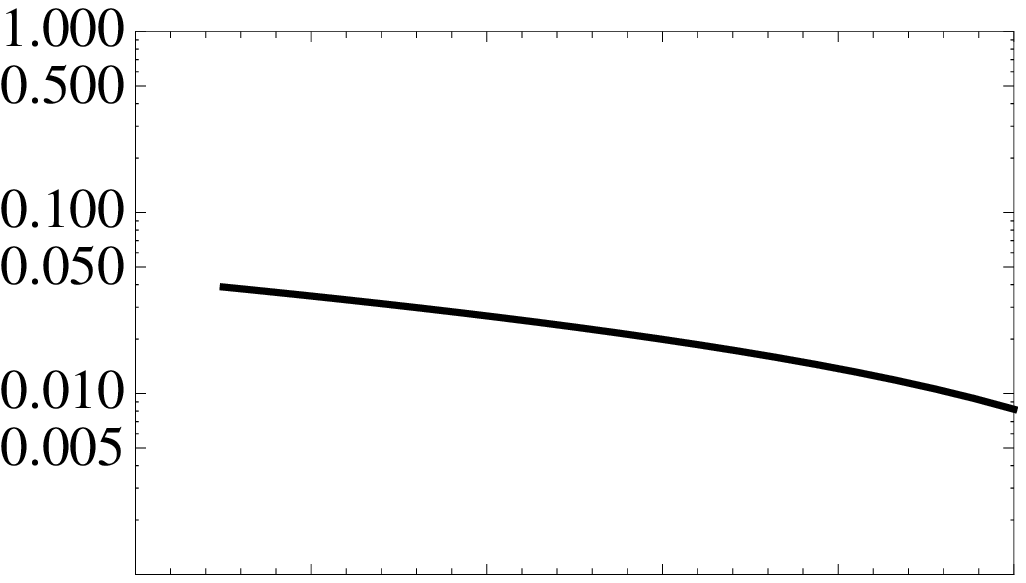}} &
\parbox[c]{130px}{\includegraphics[width=130px]{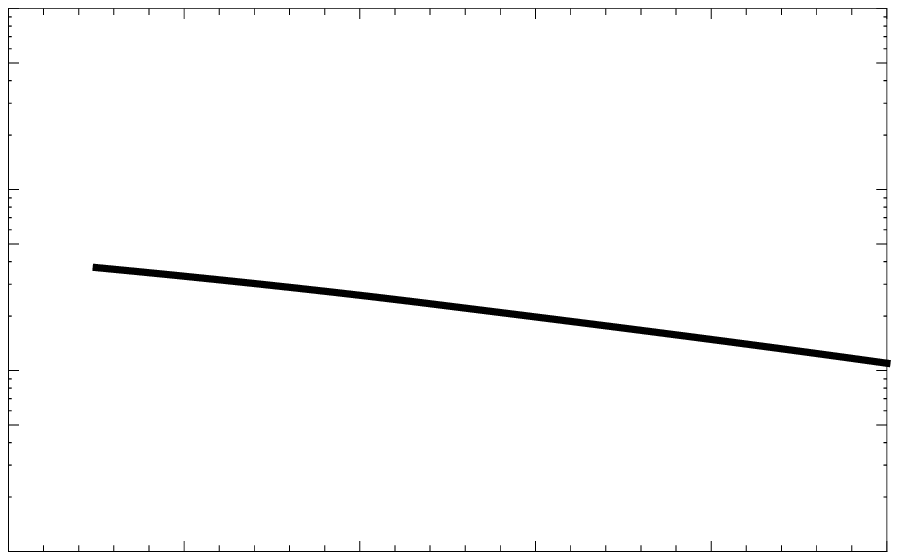}} &
\parbox[c]{130px}{\includegraphics[width=130px]{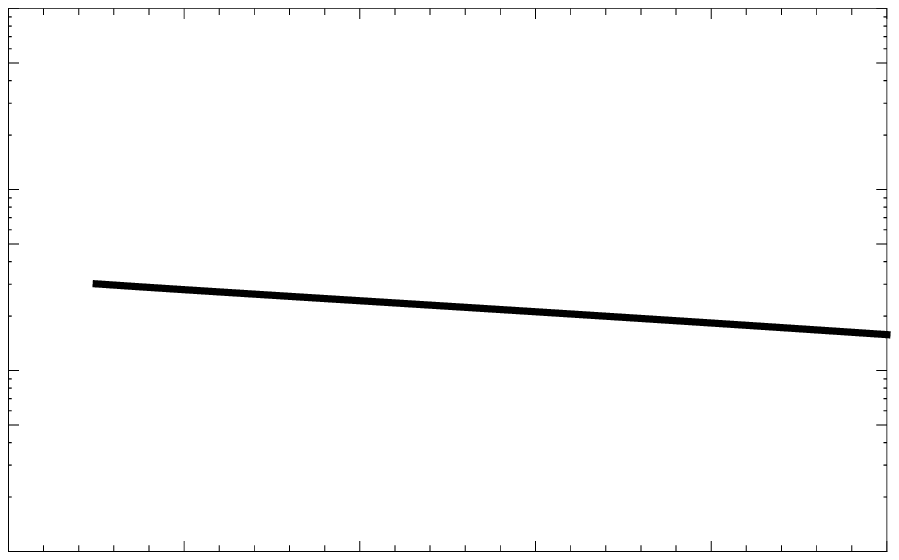}} &
\begin{turn}{-90}\hspace*{-1.4cm}\footnotesize{$\sigmavH = 450~\mathrm{km/s}$}\end{turn}\\[-22px]
\parbox[c]{130px}{\includegraphics[width=130px]{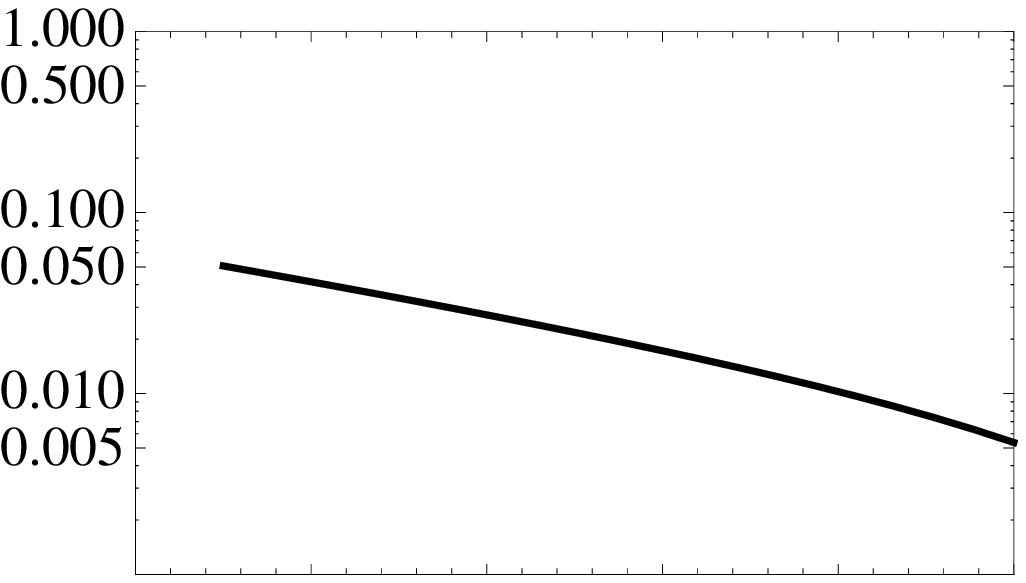}} &
\parbox[c]{130px}{\includegraphics[width=130px]{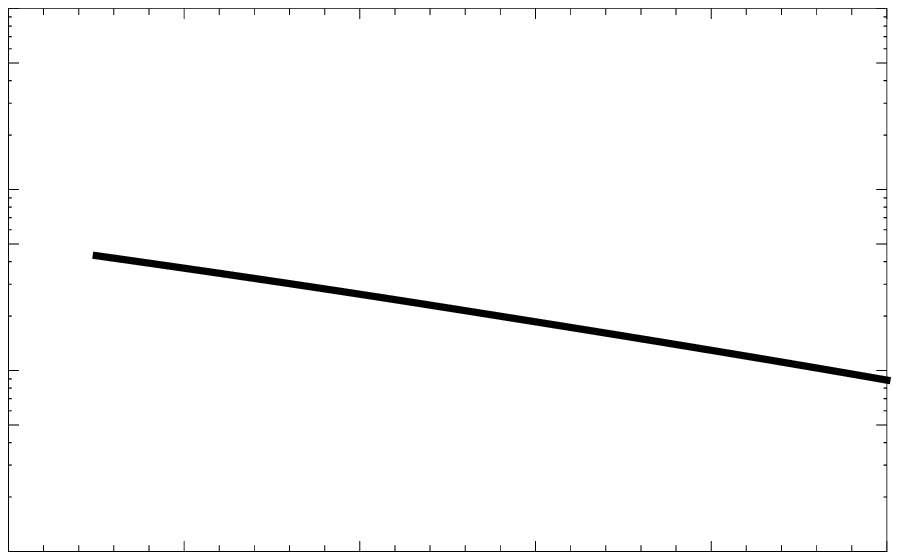}} &
\parbox[c]{130px}{\includegraphics[width=130px]{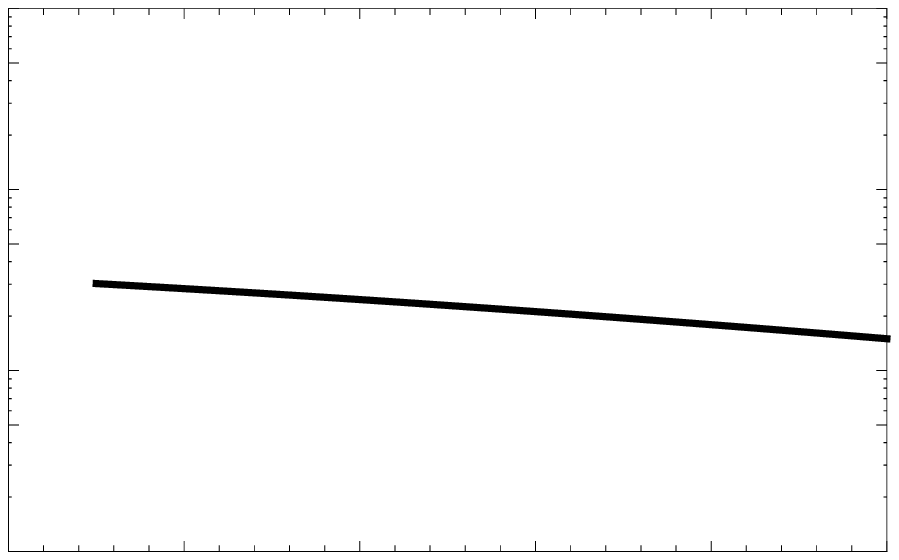}} &
\begin{turn}{-90}\hspace*{-1.4cm}\footnotesize{$\sigmavH = 250~\mathrm{km/s}$}\end{turn}\\[-22px]
\parbox[c]{130px}{\includegraphics[width=130px]{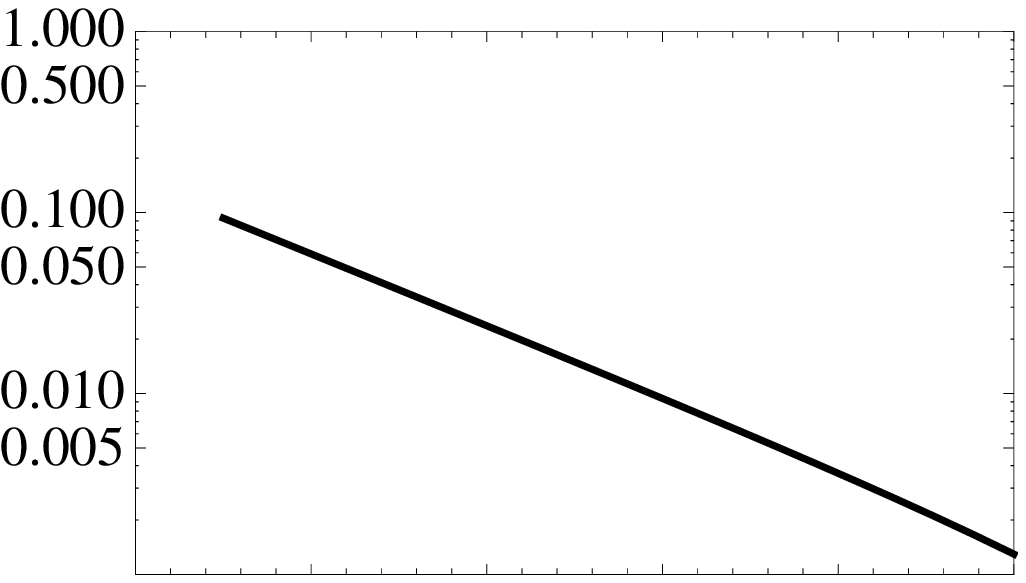}} &
\parbox[c]{130px}{\includegraphics[width=130px]{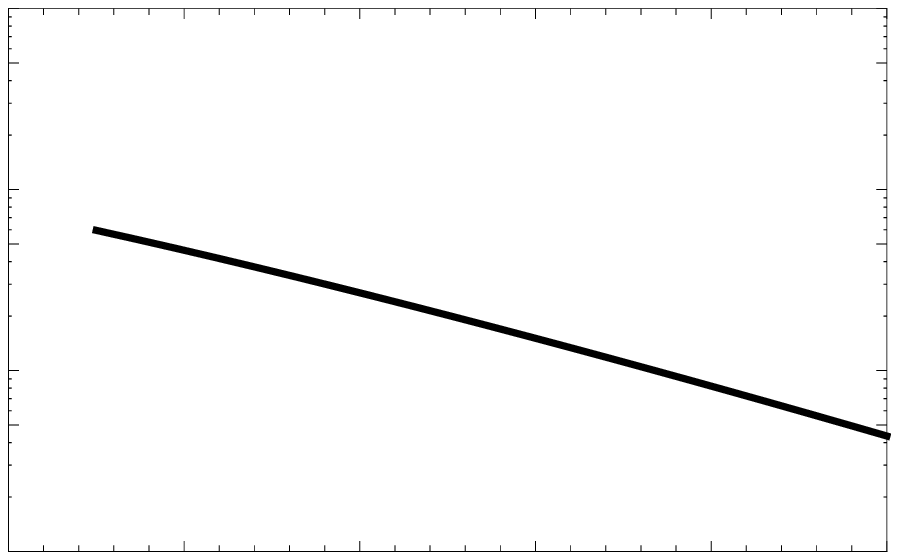}} &
\parbox[c]{130px}{\includegraphics[width=130px]{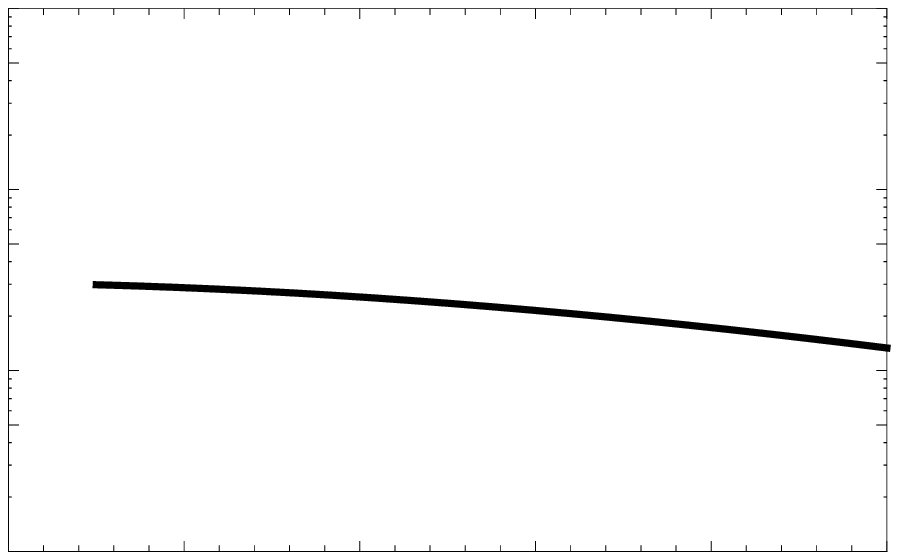}} &
\begin{turn}{-90}\hspace*{-1.4cm}\footnotesize{$\sigmavH = 155~\mathrm{km/s}$}\end{turn}\\[-22px]
\parbox[c]{130px}{\includegraphics[width=130px]{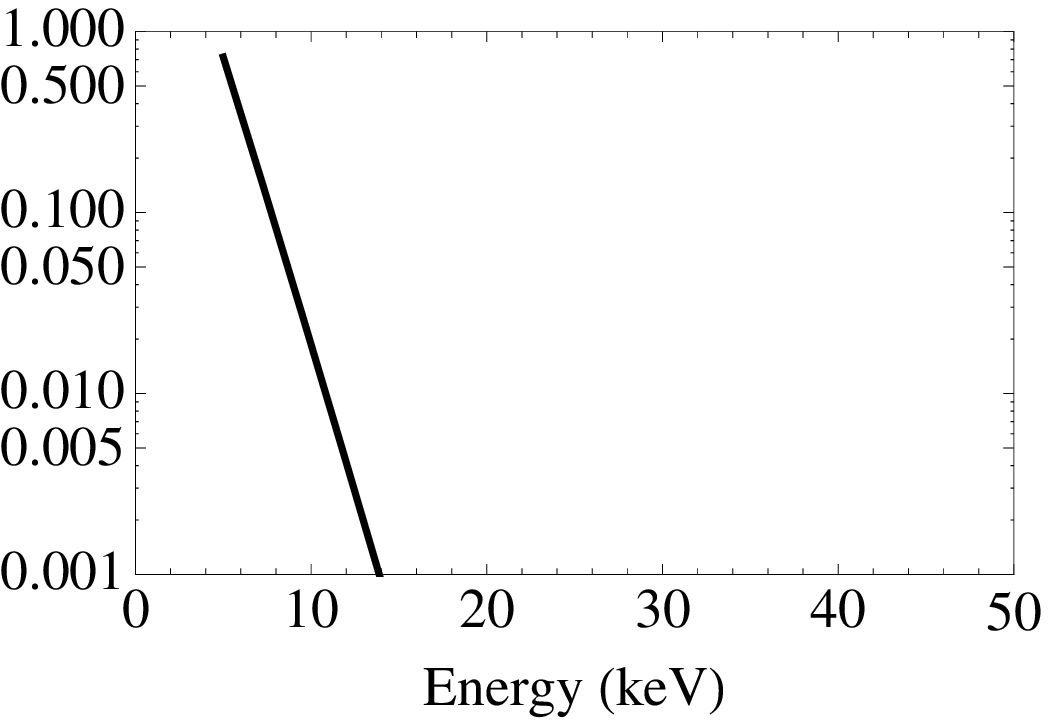}} &
\parbox[c]{130px}{\includegraphics[width=130px]{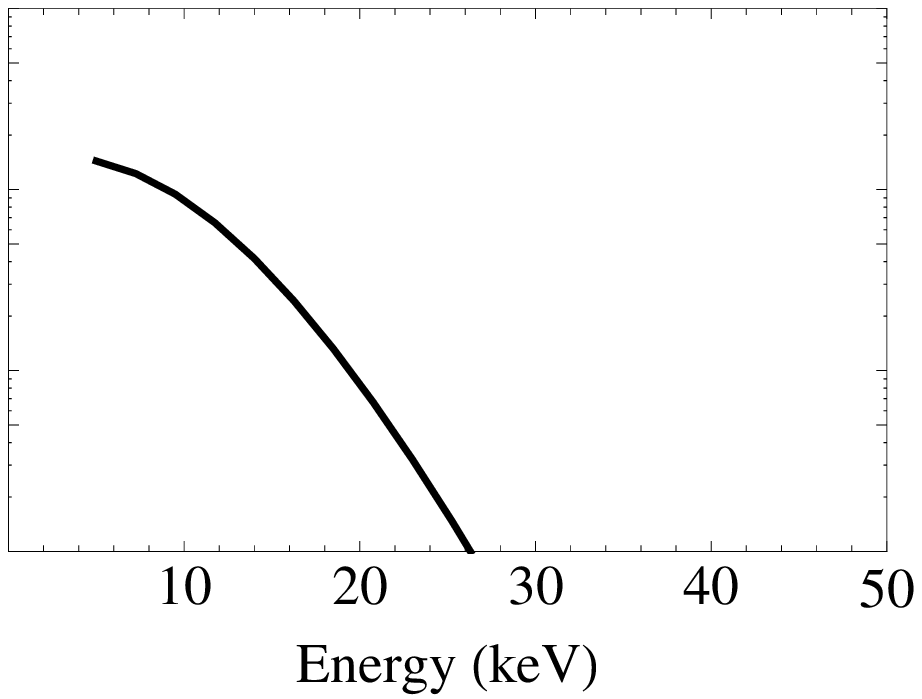}} &
\parbox[c]{130px}{\includegraphics[width=130px]{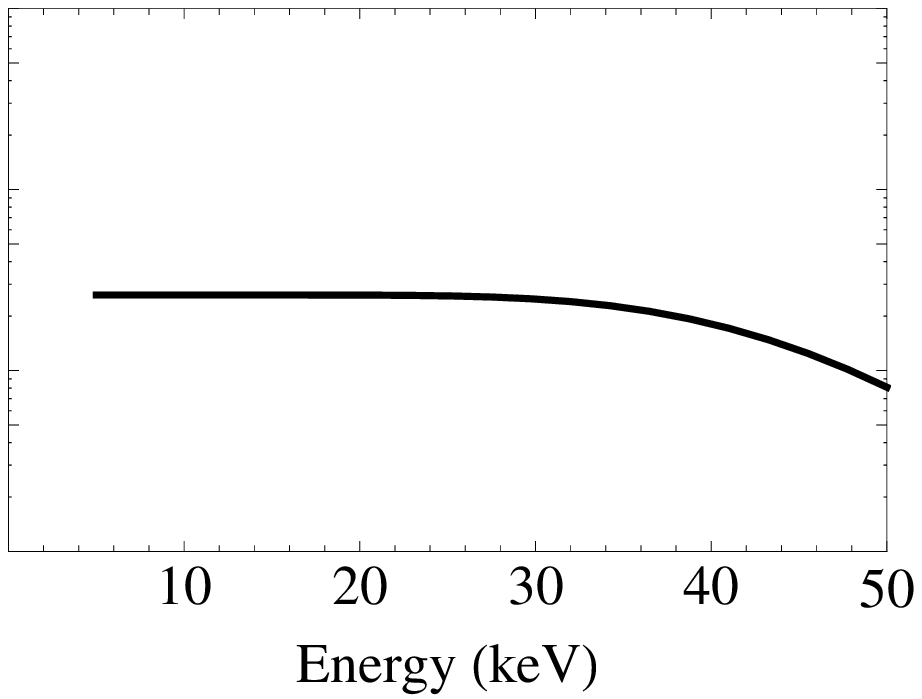}} &
\begin{turn}{-90}\hspace*{-1.4cm}\footnotesize{$\sigmavH = 50~\mathrm{km/s}$}\end{turn}\\
\end{array}$
\caption{Recoil spectra showing the energy distribution (normalized to unity) of events in the 5--50 keV energy range for a halo-only model, for the values of $\vlab$ and $\sigmavH$ indicated by the black dots in figure~\ref{fig:vonly-tri}.  The similarity of some of the spectra here leads to the parameter degeneracy in the energy-only analysis shown in that figure.}
\label{fig:vonly-grid-sp}
\end{figure}

The results of the three analyses are presented in the top row of figure~\ref{fig:vonly-tri}.  We see that the energy-only analysis yields contours for the 2D marginalized posterior probability distribution in $\vlab-\sigmavH$ space that indicate that $\vlab$ and $\sigmavH$ are anti-correlated.  This can be understood simply by noting from eq.~(\ref{eq:halospectrum}) that the dependencies of the shape of the energy spectrum on $\vlab$ and $\sigmavH$ are similar.  Increasing the value of either $\vlab$ or $\sigmavH$ results in a larger fraction of recoil events at higher energies, with both actions flattening out the exponentially-falling recoil spectrum; this behavior is illustrated in figure~\ref{fig:vonly-grid-sp}.  Thus, the free parameters $\vlab$ and $\sigmavH$ can only vary in a roughly inverse manner with each other if the observed shape of the spectrum is to be maintained.

\begin{figure}[p]
\centering
$\begin{array}{|c|c|c|c}
\multicolumn{1}{c}{\normalsize{\vlab = 50~\mathrm{km/s}}} & \multicolumn{1}{c}{\normalsize{\vlab = 220~\mathrm{km/s}}} & \multicolumn{1}{c}{\normalsize{\vlab = 450~\mathrm{km/s}}} & \\\cline{1-3}
\parbox[c]{130px}{\vspace{1px}\includegraphics[width=130px]{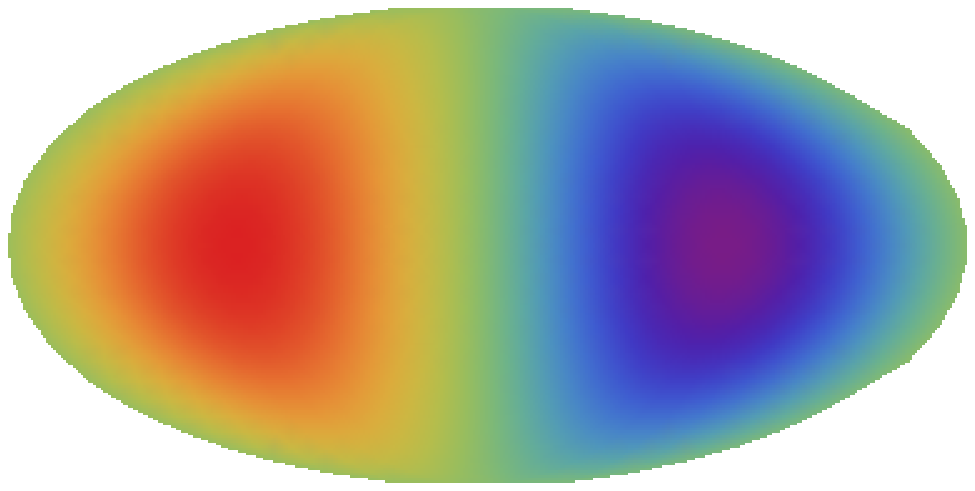}} &
\parbox[c]{130px}{\vspace{1px}\includegraphics[width=130px]{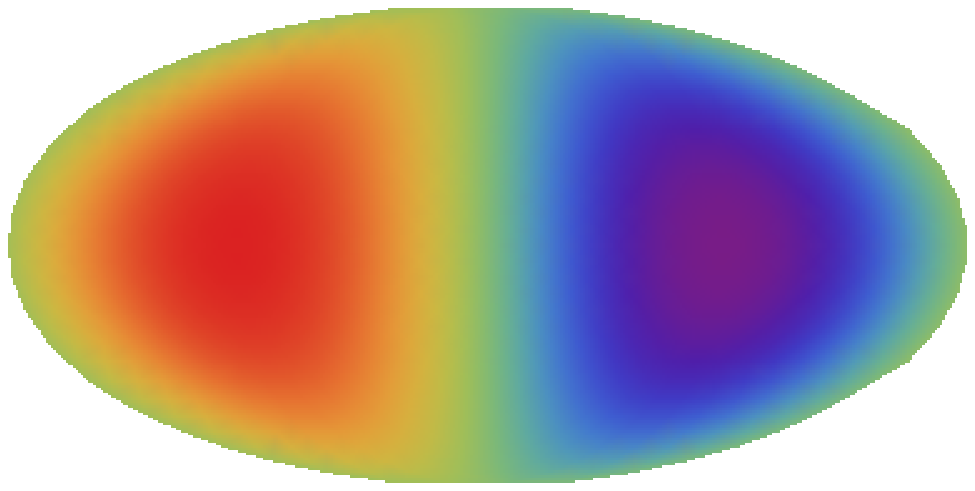}} &
\parbox[c]{130px}{\vspace{1px}\includegraphics[width=130px]{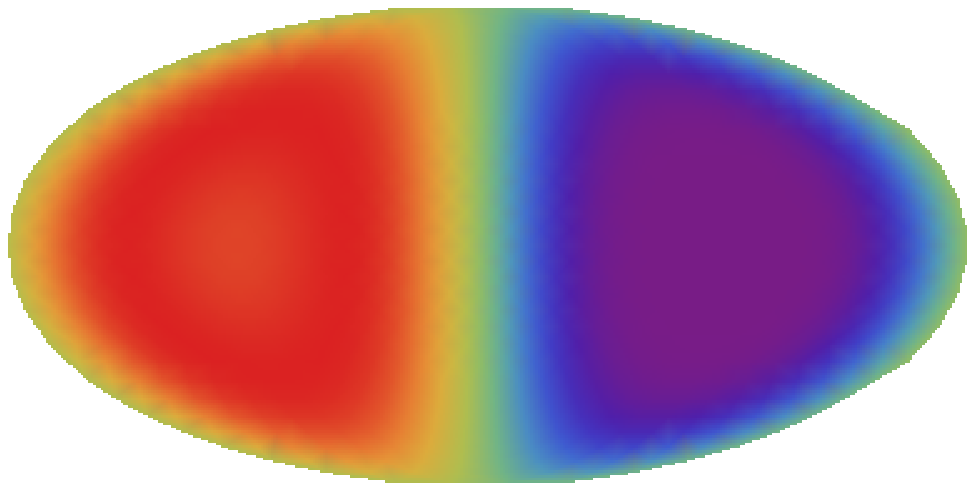}} &
\begin{turn}{-90}\hspace*{-1.15cm}\footnotesize{$\sigmavH = 450~\mathrm{km/s}$}\end{turn}\\\cline{1-3}
\parbox[c]{130px}{\vspace{1px}\includegraphics[width=130px]{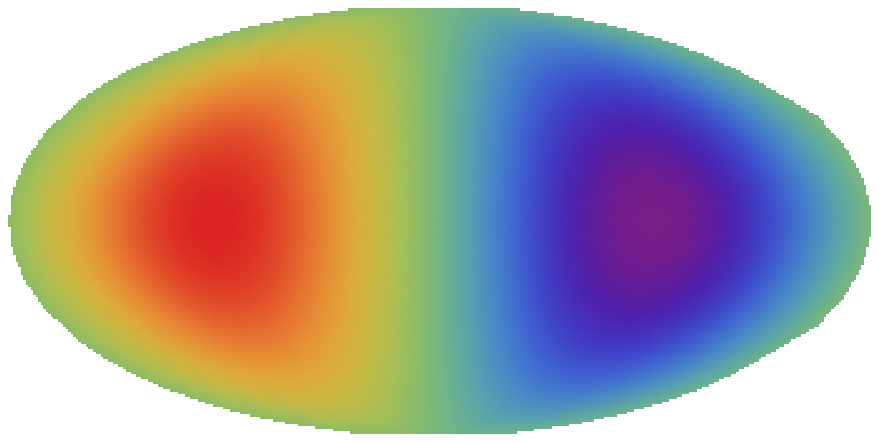}} &
\parbox[c]{130px}{\vspace{1px}\includegraphics[width=130px]{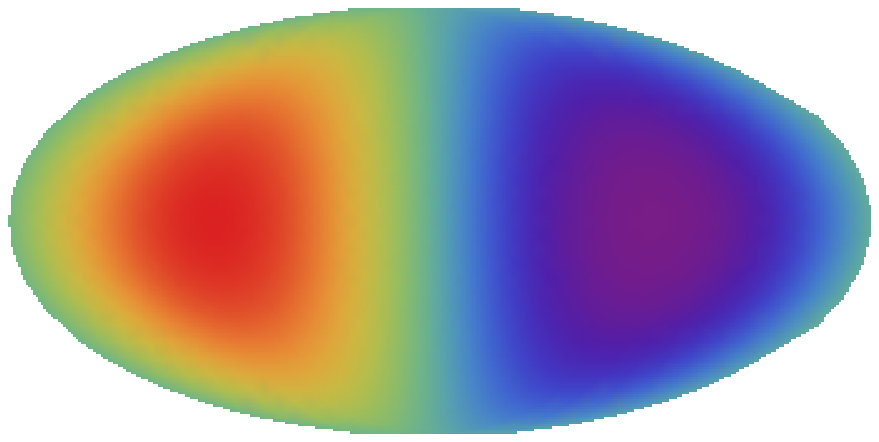}} &
\parbox[c]{130px}{\vspace{1px}\includegraphics[width=130px]{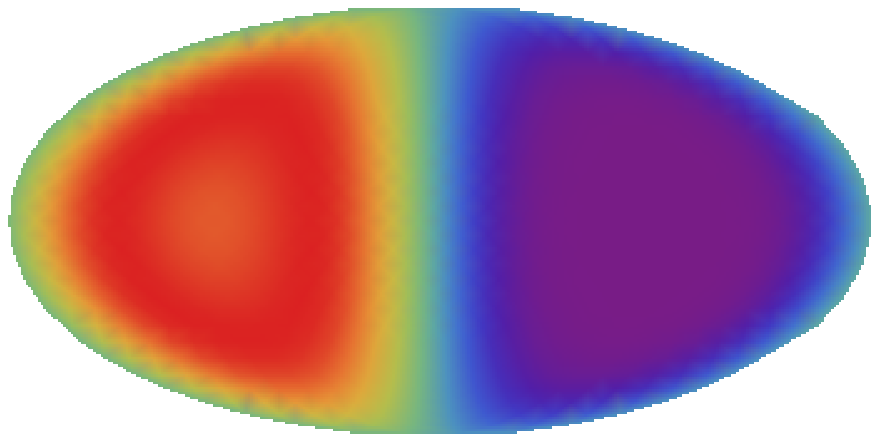}} &
\begin{turn}{-90}\hspace*{-1.15cm}\footnotesize{$\sigmavH = 250~\mathrm{km/s}$}\end{turn}\\\cline{1-3}
\parbox[c]{130px}{\vspace{1px}\includegraphics[width=130px]{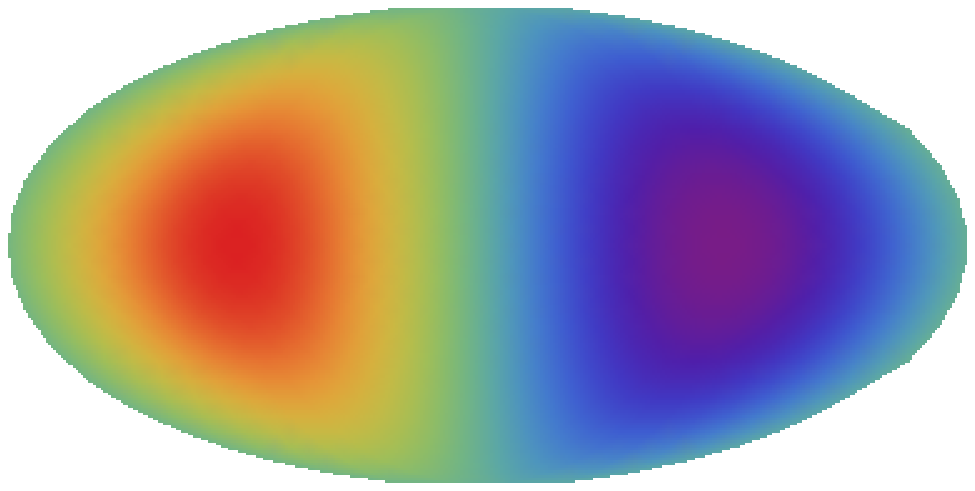}} &
\parbox[c]{130px}{\vspace{1px}\includegraphics[width=130px]{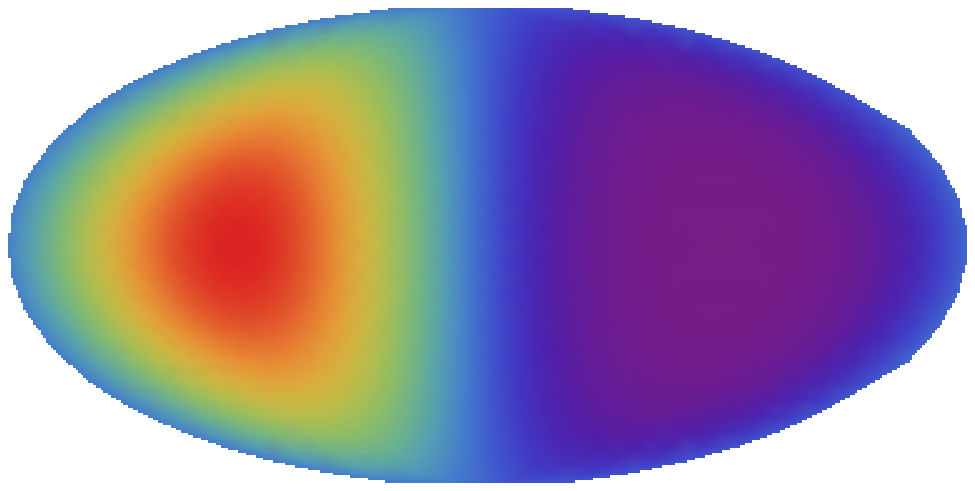}} &
\parbox[c]{130px}{\vspace{1px}\includegraphics[width=130px]{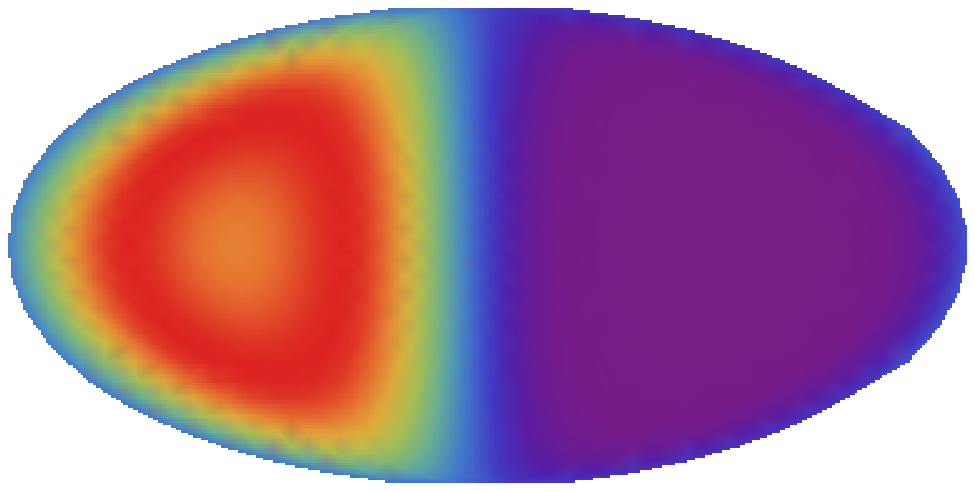}} &
\begin{turn}{-90}\hspace*{-1.15cm}\footnotesize{$\sigmavH = 155~\mathrm{km/s}$}\end{turn}\\\cline{1-3}
\parbox[c]{130px}{\vspace{1px}\includegraphics[width=130px]{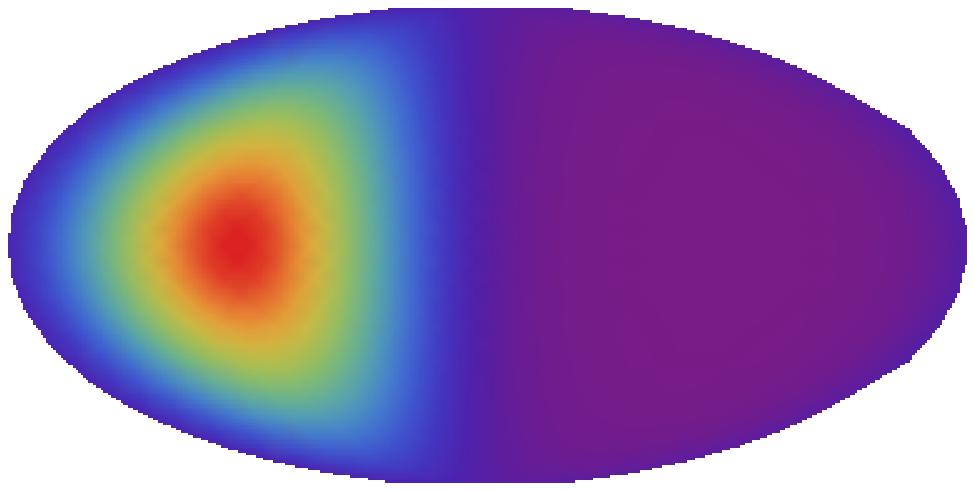}} &
\parbox[c]{130px}{\vspace{1px}\includegraphics[width=130px]{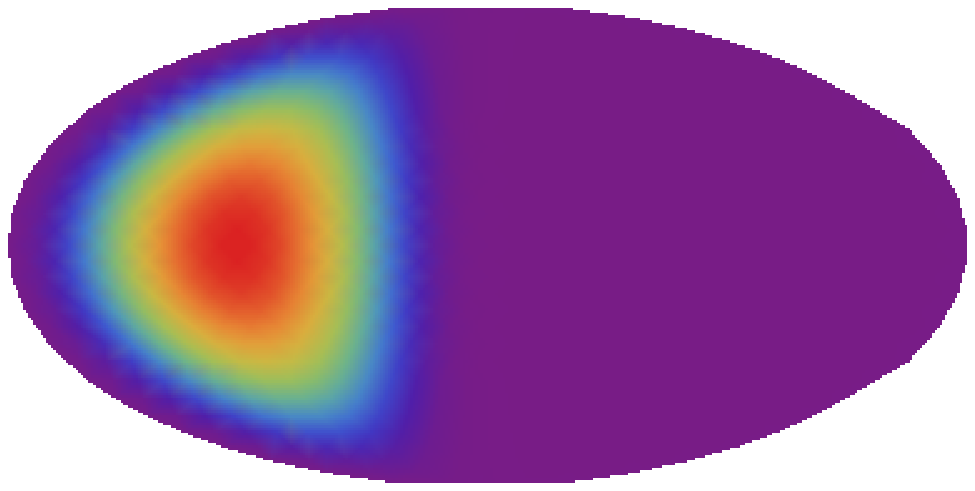}} &
\parbox[c]{130px}{\vspace{1px}\includegraphics[width=130px]{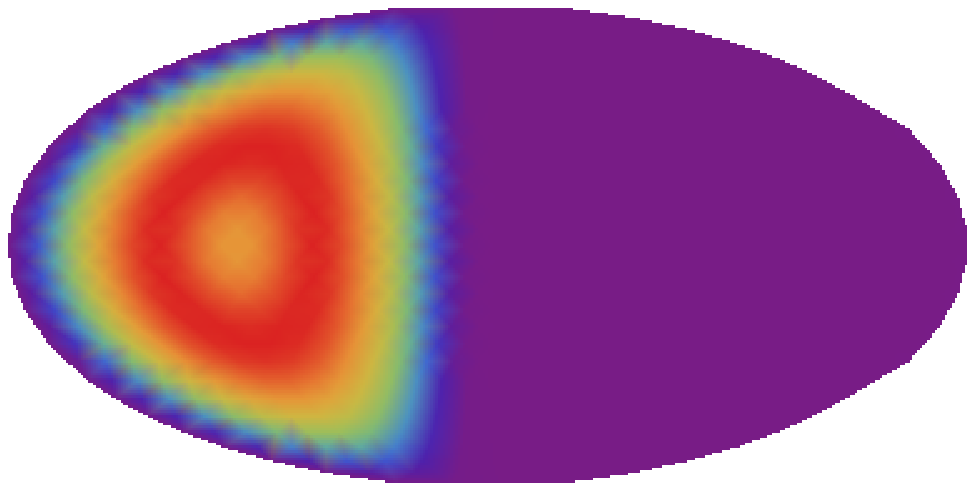}} &
\begin{turn}{-90}\hspace*{-1.1cm}\footnotesize{$\sigmavH = 50~\mathrm{km/s}$}\end{turn}\\\cline{1-3}
\end{array}$
\caption{Recoil maps showing the angular distribution (normalized to unity) of events in the 5--50 keV energy range for a halo-only model, for the values of $\vlab$ and $\sigmavH$ indicated in figure~\ref{fig:vonly-tri}.  Note the correspondence between the similarity of some of the maps and the parameter degeneracy in the direction-only analysis.  Note also that as $\vlab$ increases, a \emph{deficit} of events at the usual peak of the distribution appears, since the most energetic foward-scattering events have energies greater than 50 keV at large $\vlab$.  See also the discussion in ref. \cite{bozorgnia2011}.}
\label{fig:vonly-grid-map}
\end{figure}

On the other hand, the direction-only analysis yields contours that indicate $\vlab$ and $\sigmavH$ are correlated.  Again, this can be easily understood by considering the dependence of the directional recoil map on $\vlab$ and $\sigmavH$, illustrated in figure~\ref{fig:vonly-grid-map}.  Note that in the limit that $\vlab$ vanishes, the recoil map becomes isotropic; conversely, it is clear that increasing the value of $\vlab$ makes the map more anisotropic and asymmetric by increasing the enhancement in one hemisphere due to forward scattering by incoming particles from the ``WIMP wind.''  In contrast, the recoil map becomes more isotropic as $\sigmavH$ increases and becomes much larger than $\vlab$, since the number of incoming WIMPs arriving in the \emph{opposite} direction from the WIMP wind is then increased.  We see that $\vlab$ and $\sigmavH$ must vary in a roughly proportional manner with each other if the observed large-scale anisotropy of the recoil map is to be maintained.

Thus, the energy-only and direction-only analyses provide orthogonal sets of information on the velocity and dispersion parameters.  It is then easy to see how combining both sets of information in the direction+energy analysis yields contours that demonstrate that $\vlab$ and $\sigmavH$ are relatively uncorrelated.

We can repeat this exercise using the same recoil-event data set, this time relaxing the assumption that the WIMP mass $\mchi$ is known independently.  We now allow $\mchi$ to be an additional free parameter to be estimated, assuming a flat prior as shown in table~\ref{tab:fidparams} and sampling over the 3D $\{\mchi,\vlab,\sigmavH\}$ parameter space.  The results of the energy-only, direction-only, and direction+energy analyses are shown in the bottom row of figure~\ref{fig:vonly-tri}.  Interestingly, we see that the quality of the contours in the energy-only analysis is severely degraded compared to the case where the mass is known exactly, with a long, flat tail in the $\sigmavH$ direction appearing in the posterior probability distribution.  This can be explained in a similar manner as before; allowing for decreased values of $\mchi$ is only possible if increased values of either $\vlab$ and $\sigmavH$ compensate to fix the observed fraction of events at high energies.  However, the slope of the spectrum is slightly more sensitive to changes in $\sigmavH$ than to changes in $\vlab$, as can be shown by examining eq.~(\ref{eq:halospectrum}) (in particular, by considering the absolute magnitude of the derivatives of $dR/dE$ with respect to $\vlab$ and $\sigmavH$ in the relevant regions of parameter space).  Thus, small values of $\mchi$ are more easily compensated for by increasing $\sigmavH$, so a long tail appears in the $\sigmavH$ direction.

It is notable that the analyses incorporating directional information are relatively insensitive to the lack of prior knowledge of the WIMP mass.  This again demonstrates the power and robustness of combining directional and energy information to fix parameters of the velocity distribution.

\subsubsection{6-parameter Analyses} \label{subsec:6param}

\begin{figure}[p]
\centering
$\begin{array}{c}
\parbox[c]{3in}{\includegraphics[width=3in]{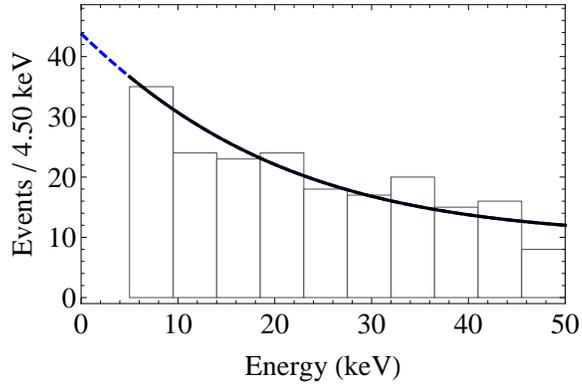}}
\end{array}$
\caption{Simulated recoil spectrum for the halo-only 6-parameter analysis in section~\ref{subsec:6param}, with binned signal and background events.  The fiducial halo-only spectrum with a flat background is also plotted.}
\label{fig:h-sp}
\end{figure}

\begin{figure}[b]
\centering
$\begin{array}{c}
\parbox[c]{6in}{\includegraphics[width=6in]{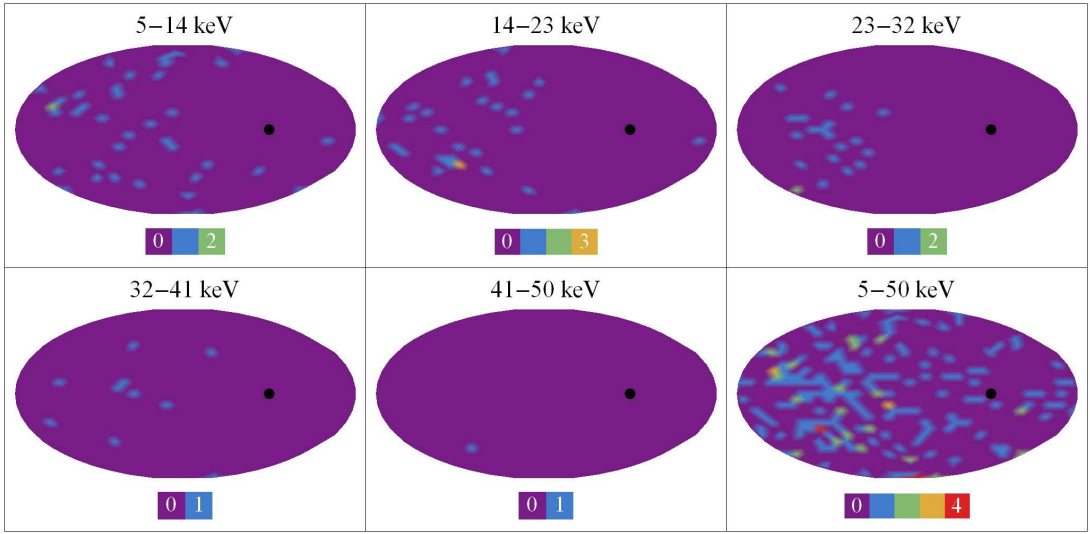}}
\end{array}$
\caption{Simulated recoil maps for the halo-only 6-parameter analysis in section~\ref{subsec:6param}.  Maps of the signal events due to WIMP-induced recoils in 5 different energy bins are shown, along with a map including both signal and background events.  Note that the clustering of events tightens as their energy increases because of the kinematics of the scattering process.}
\label{fig:h-map}
\end{figure}

We now proceed to perform likelihood analyses over the full 6D standard-halo-model parameter space of ${\{\mchi, \lambda, \vlab, \llab, \blab, \sigmavH\}}$.  Using the same fiducial values for the velocity-distribution parameters as before, we simulate a total of $\Ntot = 200$ events and take $\lambda = 0.5$, corresponding to $\Nsig = 100$ and $\Nbg = 100$.  For a MIMAC-like experiment with 10 kg of target mass, assuming a WIMP mass $\mchi = 50~\textrm{GeV}$, a typical local density $\rho_0 \approx 0.3~\mathrm{GeV/cm^3}$, and a spin-dependent WIMP-nucleon cross section of $\sigma_{p,n} \approx 10^{-3}~\textrm{pb}$ (consistent with current observational limits from both direct-detection and neutrino experiments \cite{tanaka2011,abbasi2011,picasso2012}), this number of signal events roughly corresponds to a 5-month observation period.  The number of background events then corresponds to the assumption of a relatively high background event rate of $\sim\!\!23$/kg/yr (in comparison, it may be reasonable to expect background event rates as low as 10/kg/yr \cite{billard2011}).  We again take $\Npix = 768$, and bin events into $\Nbins = 10$ energy bins.  The recoil spectrum and binned recoil sky maps for the simulated data are shown in figures~\ref{fig:h-sp}~and~\ref{fig:h-map}, respectively.

\begin{figure}[p]
\centering
\includegraphics[width=6in]{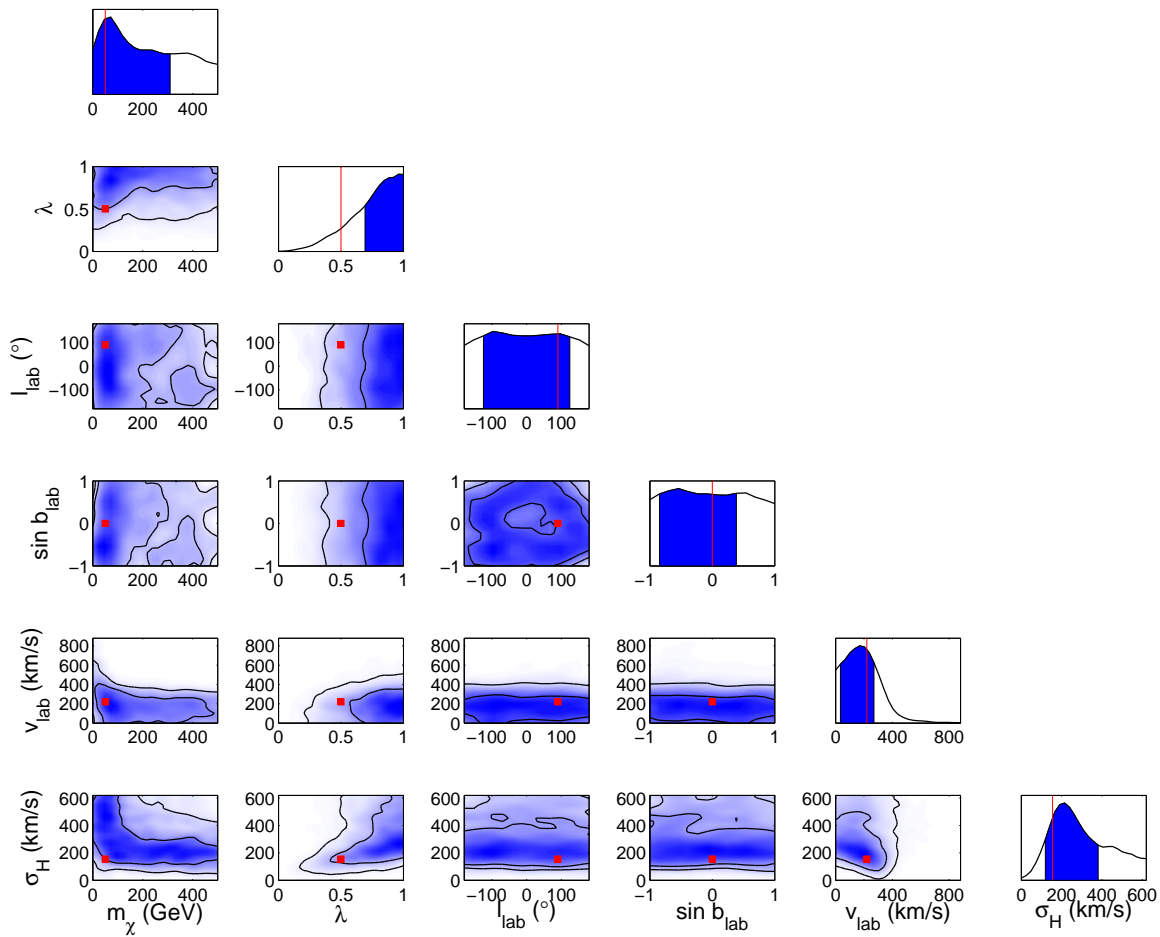}
\caption{Triangle plot showing 1D and 2D posterior probability distributions over the full prior ranges, for the halo-only 6-parameter analysis using only energy information in section~\ref{subsec:6param}. Red lines and square markers indicate the fiducial parameter values used in simulating the data.   On the 1D plots, $68\%$ minimum credible intervals are shaded in blue.  Parameter estimation is relatively poor.}
\label{fig:h-enonly-tri}
\end{figure}

\begin{figure}[p]
\centering
\includegraphics[width=6in]{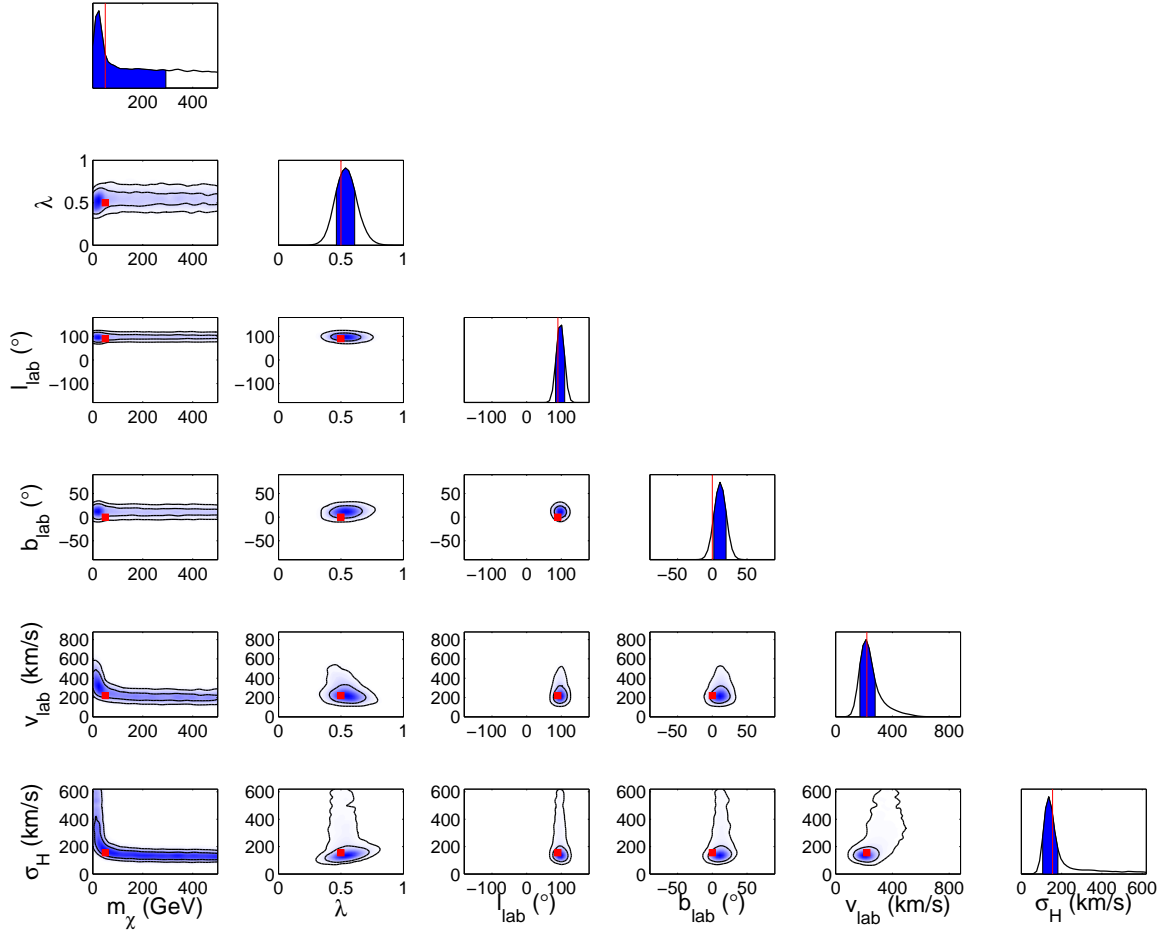}
\caption{The same as figure~\ref{fig:h-enonly-tri}, but for the halo-only 6-parameter analysis using direction+energy information in section~\ref{subsec:6param}.  Parameter estimation is much improved over the energy-only analysis.}
\label{fig:h-diren-tri}
\end{figure}

We shall perform both an energy-only ($\Npix = 1$) analysis  and a direction+energy analysis.  The results of the parameter estimation are shown in the triangle plots of the 1D and 2D marginalized posterior probability distributions in figures~\ref{fig:h-enonly-tri}~and~\ref{fig:h-diren-tri}.  Note that the $68\%$ MCIs for each parameter are plotted there, and are also given in table~\ref{tab:fidparams} along with the 1D marginal posterior modes.

From the triangle plots, it is clear that the analysis incorporating the directional information is able to recover the parameters with greater fidelity than the energy-only analysis.  Not only does this information allow for the direction $(\llab,\blab)$ of the lab frame to be recovered quite accurately, it also allows a rough measurement of the background-rejection power $\lambda$ from the data itself.  That is to say, the directional information indeed allows the isotropic background component to be separated from the anisotropic signal component.  In contrast, the energy-only analysis is unable to recover $\lambda$ correctly.  As we saw above, flattening of the exponentially-falling spectrum may be caused by variations in the other parameters, and cannot be easily separated from the introduction of a truly flat background spectral component using only the energy information (at least, not without improved statistics from a larger number of events).

Note also that the estimate of $\mchi$ is quite poor for both analyses, with a long tail extending to large values of $\mchi$.  This partially stems from the fact that we are only using information from the \emph{distribution} of the events, which only has weak dependence on $\mchi$ as $\mchi$ increases, as discussed previously.  Since there is some additional dependence of the amplitude of the signal on $\mchi$, as can be seen from eq.~(\ref{eq:spectrum}), making the aforementioned assumptions about the additional amplitude-fixing parameters $\rho_0$ and $\sigma_N$ can greatly improve these estimates of $\mchi$.

Finally, from examination of the plots for $\vlab$ and $\sigmavH$ in figures~\ref{fig:h-enonly-tri}~and~\ref{fig:h-diren-tri}, we see that the intuitive results about the ability to constrain these parameters from the simple 2-parameter and 3-parameter analyses are basically borne out even in the full 6-parameter analysis.  However, we note that the presence of a flat isotropic background at this level does degrade the ability to pin down $\sigmavH$, even with directional information, as is evident from the long tail in the marginalized posterior probability distribution for $\sigmavH$.  This tail may likewise be reduced by improved statistics or better background rejection.

To summarize, for the standard halo model, 100 signal events (which might reasonably be observed in a 5-month period) yield a good measurement of the direction of the LSR, and allows rough constraints to be placed on the velocity and dispersion of the Galactic halo, the presence of a non-negligible background notwithstanding.  Of course, the results may be improved simply by lengthening the observation period; a 3-year observation period resulting in a 30-kg-yr exposure might be expected.  In such a period, for the fiducial parameter values we have assumed for the standard halo model and WIMP properties, one would expect $\sim\!\!650$ signal events within the energy-sensitivity range 5--50 keV (and $\sim\!\!900$ total events down to zero threshold).  It is then interesting to ask whether or not any interesting structure in the local dark-matter velocity distribution, such as cold streams or disk components, may be detected with a comparable number of events.  We shall proceed to investigate this question in the subsections to follow.

\subsection{Halo+stream Model} \label{subsec:halostream}

Motivated by the results of the simulations mentioned in section~\ref{sec:intro}, we next consider a model with a local dark-matter stream in addition to the dark-matter halo.  We shall assume that the velocity distribution in the Galactic frame of a stream component is given by
\begin{equation}
\fg^\textrm{S}(\bold{\vg}) = \fgTM(\bold{\vg-\vSg}; \sigmavS,\vesc)\,,
\end{equation}
where the velocity vector of the stream in the Galactic frame is $\bold{\vSg}$ (and may be indicated by its magnitude $\vSg$ and its direction $(\lS,\bS)$ in Galactic coordinates) and the velocity dispersion $\sigmavS$ is small for a cold tidal stream.  We further assume that the dark-matter particles in the stream consist of a fraction $\AS$ of the total particles in both the halo and the stream locally.  Using the linearity of the Radon transform and eq.~(\ref{eq:radonprop}), we find
\begin{equation}
\widehat{f^{\textrm{H+S}}}(\vq,\widehat{\bold{q}}) = (1-\AS) \widehat{\fgTM}(\vq + \bold{\vlab} \cdot \bold{\widehat{q}}, \bold{\widehat{q}}; \sigmavH,\vesc) + \AS \widehat{\fgTM}(\vq + (\bold{\vlab-\vSg})\cdot \bold{\widehat{q}}, \bold{\widehat{q}}; \sigmavS,\vesc)\,,
\end{equation}
which yields the directional recoil spectrum via eq.~(\ref{eq:spectrum}).

We shall assume the fiducial values of the parameters ${\{\mchi, \vlab, \llab, \blab, \sigmavH\}}$ determining the halo component of the distribution are identical to those used in the halo-only model above.  For simplicity, we shall further assume that the parameter $\vlab = \vLSR = 220~\textrm{km/s}$ is known exactly, so that the halo component is specified by 4 free parameters.  This is done simply to differentiate between the two Maxwellian components of the velocity distribution within the likelihood analysis, which would otherwise be arbitrarily assigned.  By fixing $\vlab$, we can identify the ``halo'' as the Maxwellian component that moves with mean velocity $\vlab$ with respect to the lab frame, while the ``stream'' is identified as the secondary component.  For the 5 stream parameters, we shall take the fiducial values $\{\AS = 0.1$, $\lS = 65^{\circ}$, $\bS = 25^{\circ}$, $\vSg = 510~\textrm{km/s}$, $\sigmavS = 10~\textrm{km/s}\}$.  Including $\lambda$, the halo+stream model is specified by 10 parameters in total.

For this halo+stream model, we shall perform only a direction+energy analysis, assuming we have observed an number of events comparable to that expected in the baseline halo-only scenario with a 30 kg-yr total exposure.  For the experimental parameters, we adopt $\{\Npix = 768$, $\Nbins = 20$, $\Ntot = 950$, $\lambda = 650/950 \approx 0.684\}$.  This yields $\Nsig = 650$ and $\Nbg = 300$, roughly corresponding to an exposure of 28.9-kg-yr and background rate of 10.4/kg/yr.  These values are close to those in the baseline model, since the $10\%$ stream component is merely a small perturbation of the standard halo-only model.

\begin{figure}[p]
\centering
$\begin{array}{c}
\parbox[c]{3in}{\includegraphics[width=3in]{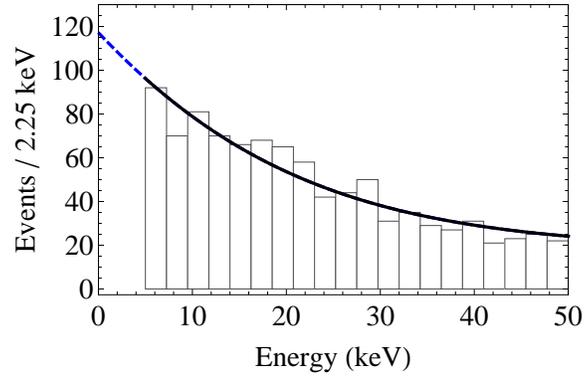}}
\end{array}$
\caption{Simulated recoil spectrum for the halo+stream analysis in section~\ref{subsec:halostream}.  The fiducial halo+stream spectrum with a flat background is also plotted.}
\label{fig:hs-sp}
\end{figure}

\begin{figure}[b]
\centering
$\begin{array}{c}
\parbox[c]{6in}{\includegraphics[width=6in]{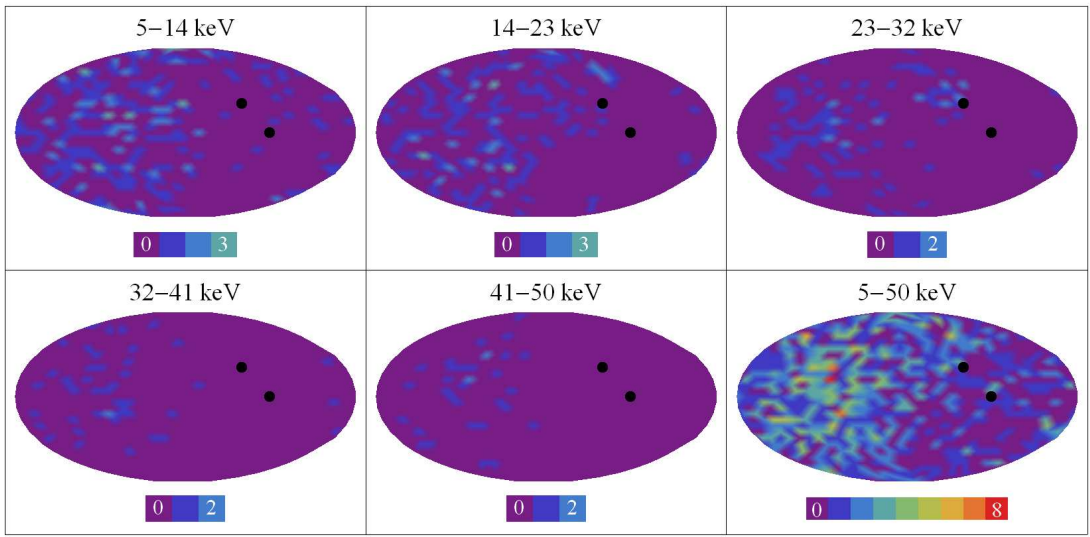}}
\end{array}$
\caption{Simulated recoil maps for the halo+stream analysis in section~\ref{subsec:halostream}, as in figure~\ref{fig:h-map}.  Black dots indicate the direction of the LSR and the stream.  The cluster of stream events is most easily seen in the top-right panel, although it is slightly visible as a wider ring-like feature at lower energies.}
\label{fig:hs-map}
\end{figure}

\begin{sidewaysfigure}
\centering
\includegraphics[height=6in]{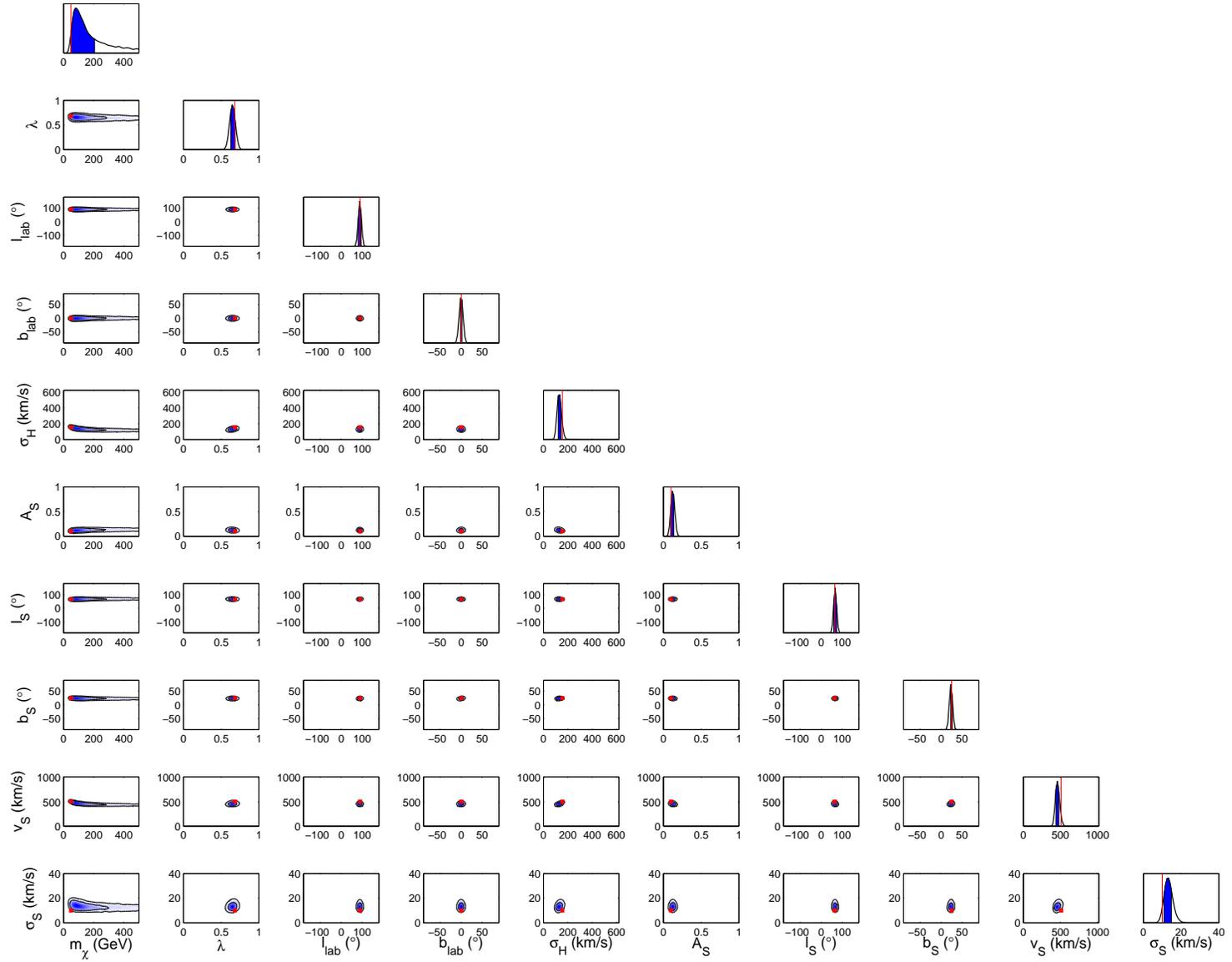}
\caption{Triangle plot for the halo+stream analysis in section~\ref{subsec:halostream}, using direction+energy information.  Parameter estimation is relatively good.}
\label{fig:hs-tri}
\end{sidewaysfigure}

The recoil spectrum and sky maps for the simulated data are shown in figures~\ref{fig:hs-sp}~and~\ref{fig:hs-map}, respectively.  The triangle plot of the posterior probability distributions resulting from the direction+energy analysis is shown in figure~\ref{fig:hs-tri}, and the posterior modes and MCIs are given in table~\ref{tab:fidparams}.  The quality of parameter estimation is quite good; with the exception of $\mchi$, all the parameters are recovered accurately without bias and with a low degree of correlation.

This seems to suggest that streams may be detectable with a MIMAC-like experiment.  However, let us note that we have adopted a somewhat unrealistically high stream fraction $\AS$ here (solely for illustrative purposes, e.g., to make visible  in the recoil maps in figure~\ref{fig:hs-map} the feature arising from the stream).  More realistically, simulations suggest that typical stream fractions are more likely at the $\sim\!\!1\%$ level, and furthermore, that the probability that a tidal stream dominates the local neighborhood of the Sun (averaged on kpc scales) is less than $1\%$ \cite{vogelsberger2009,maciejewski2011}.  Nevertheless, our analysis here serves as a proof of principle for the possibility of detecting structure in the velocity distribution beyond the standard halo model and using directional information to constrain parameters.  Furthermore, for a more realistic stream fraction of $\AS \sim 1\%$, one might expect only a $\sim\!\!\sqrt{10}$ reduction in statistical power when constraining the stream parameters.  Finally, this result suggests that one would expect comparable -- or even improved -- statistical power when constraining warm debris flows, which may have similarly large mean velocities but are thought to compose a more significant fraction (tens of percent, becoming dominant at large velocities) of the local dark-matter density.

\subsection{Halo+disk model} \label{subsec:halodisk}

Finally, we consider a Galactic model with a dark-matter-disk component in addition to the dark-matter halo \cite{read2008,read2009}.  We assume that disk rotates such that local particles in the disk move with some average velocity $\bold{\vDg}$ with respect to the halo/Galactic frame; this velocity may be specified by its magnitude $\vDg$ and its direction $(\lD, \bD)$ in Galactic coordinates.  We shall take $\bold{\vDg}$ to be parallel to $\bold{\vLSR}$, so that these particles lag the LSR by $\vlag = \vDg - \vLSR$; typical values from simulations are $\vDg \approx 170$ km/s and $\vlag \approx -50$ km/s, such that the dark disk rotates more slowly than the stellar disk.  We again assume that the local velocity distribution of the disk component is also a truncated Maxwellian with dispersion $\sigmavD$ and that particles in the disk compose a fraction $\AD$ of the local particles, so that the disk velocity distribution is identical to that of halo+stream model for $\{\AS,\vSg,\sigmavS\} \to \{\AD,\vDg,\sigmavD\}$.

We again use the fiducial values of the halo component parameters, assuming $\vlab = \vLSR = 220~\mathrm{km/s}$ is known as before.  For the disk parameters, we shall take the fiducial values $\{\AD = 0.5$, $\lD = 90^{\circ}$, $\bD = 0^{\circ}$, $\vDg = 170~\textrm{km/s}$, $\sigmavD = 100~\textrm{km/s}\}$.

It is clear that detecting nuclear recoils from collisions with low-velocity WIMPs in the disk component may be challenging, since the average lab-frame speed $\vDg = 170~\textrm{km/s}$ for such WIMPs is close to the $\sim\!\!150~\textrm{km/s}$ velocity threshold (corresponding to the 5 keV energy threshold, with our fiducial values for $\mN$ and $\mchi$).   Therefore, to investigate the effect of the energy threshold on the parameter estimation, we shall perform two direction+energy analyses: the first with a zero-energy threshold, and the second with a 5-keV threshold as before.

Again, we would like to evaluate the statistical power of our halo+disk analysis, assuming we have observed an number of events comparable to that expected in the baseline halo-only scenario with a 30-kg-yr exposure.  As mentioned previously, this baseline scenario would yield $\sim\!\!900$ signal events in the 0--50 keV range, along with 300 background events (assuming the expected rate of 10/kg/yr).  We therefore adopt we adopt $\{\Npix = 768$, $\Nbins = 40$, $\Ntot = 1200$, $\lambda = 0.75\}$ for the first zero-threshold analysis.  For the second analysis, we then make a cut of all events below the 5-keV threshold, resulting in $\{\Nbins = 36$, $\Ntot = 813$, $\lambda = 541/813 \approx 0.665\}$.  Worth mentioning here is the fact that simulations show that the presence of a disk component \emph{enhances} the local dark-matter density over that from the halo alone, so that $\rho_0 \to \rho_0/(1-\AD)$.  This correspondingly decreases the amount of exposure needed to reach the baseline of 900 signal events, and also allows for a larger background rate.

\begin{figure}[p]
\centering
$\begin{array}{c}
\parbox[c]{3in}{\includegraphics[width=3in]{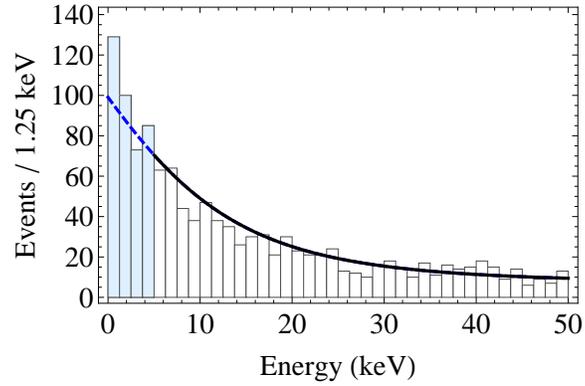}}
\end{array}$
\caption{Simulated recoil spectrum for the halo+disk analyses in section~\ref{subsec:halodisk}.  The fiducial halo+disk spectrum with a flat background is also plotted.  The shaded light-blue bars indicate the binned events below the 5-keV energy threshold included in the second 0--50 keV analysis.}
\label{fig:hd-sp}
\end{figure}

\begin{figure}[b]
\centering
$\begin{array}{c}
\parbox[c]{6in}{\includegraphics[width=6in]{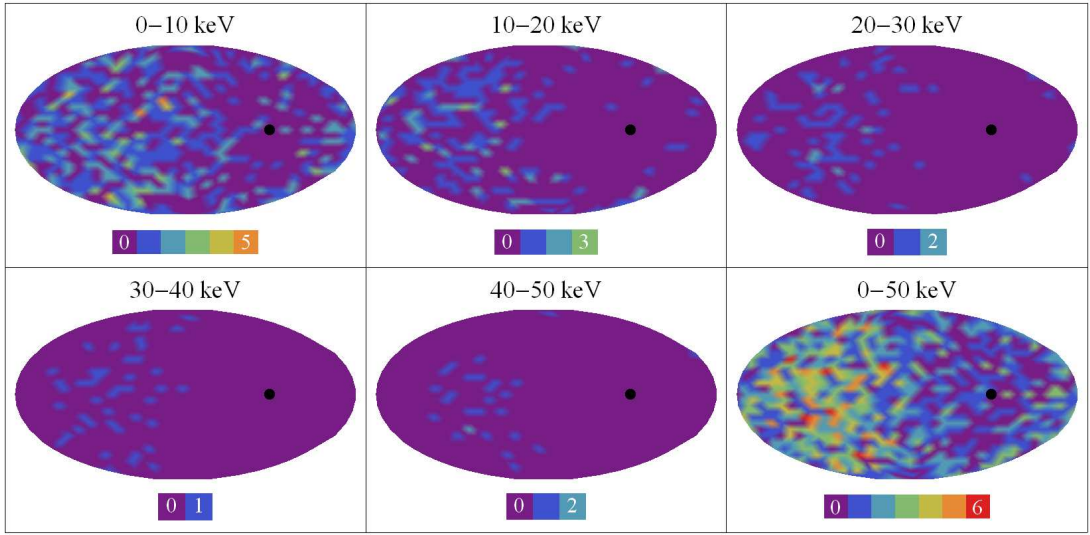}}
\end{array}$
\caption{Simulated recoil maps for the halo+disk analyses in section~\ref{subsec:halodisk}, as in figure~\ref{fig:h-map}.  Black dots indicate the common direction of the LSR and the disk.   Events over the full 0--50 keV energy range are shown.}
\label{fig:hd-map}
\end{figure}

\begin{sidewaysfigure}
\centering
\includegraphics[height=6in]{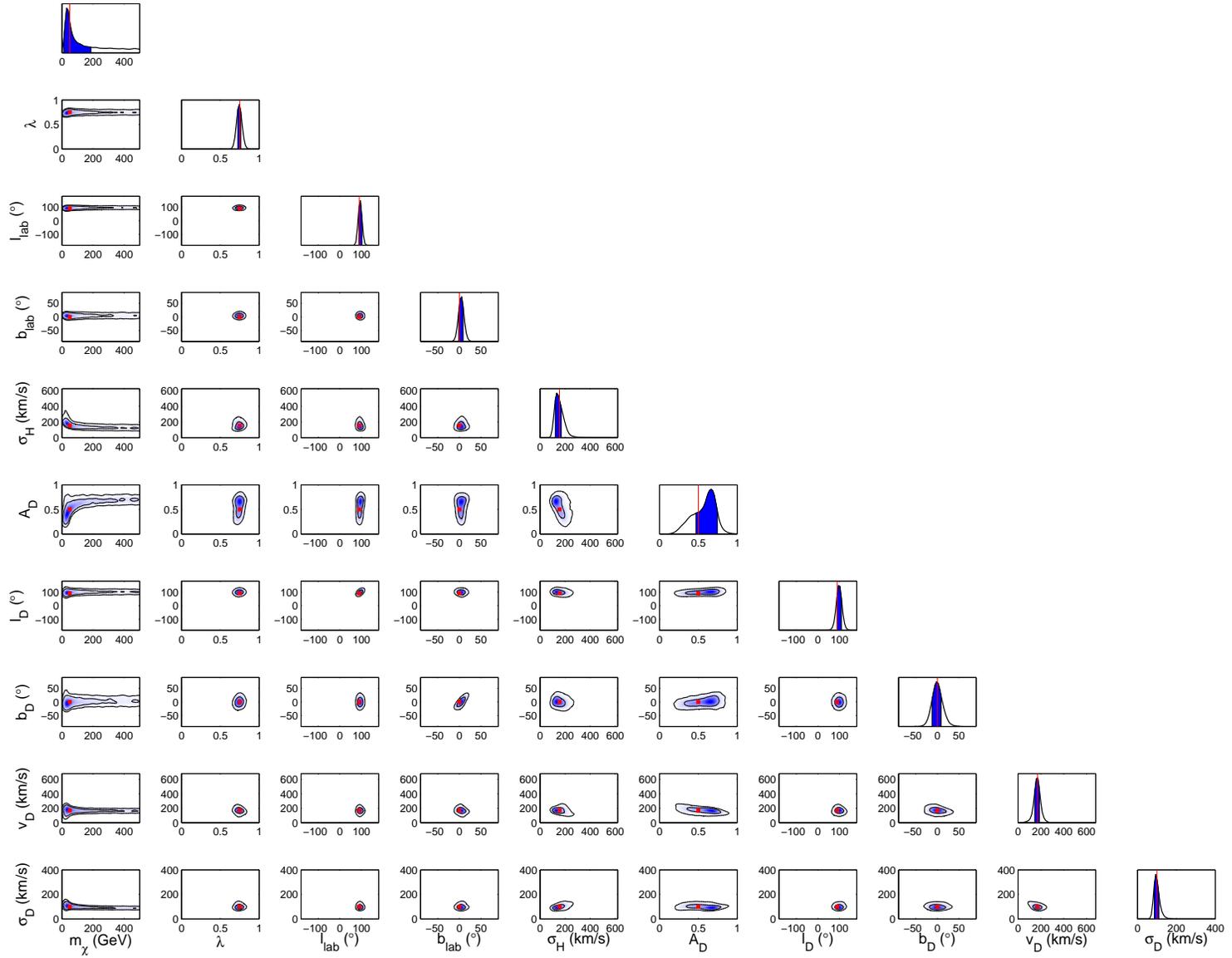}
\caption{Triangle plot for the 0--50 keV halo+disk analysis in section~\ref{subsec:halodisk}.  Parameter estimation is fair, but some of the disk parameters are recovered with less accuracy.}
\label{fig:hd-tri}
\end{sidewaysfigure}

\begin{sidewaysfigure}
\centering
\includegraphics[height=6in]{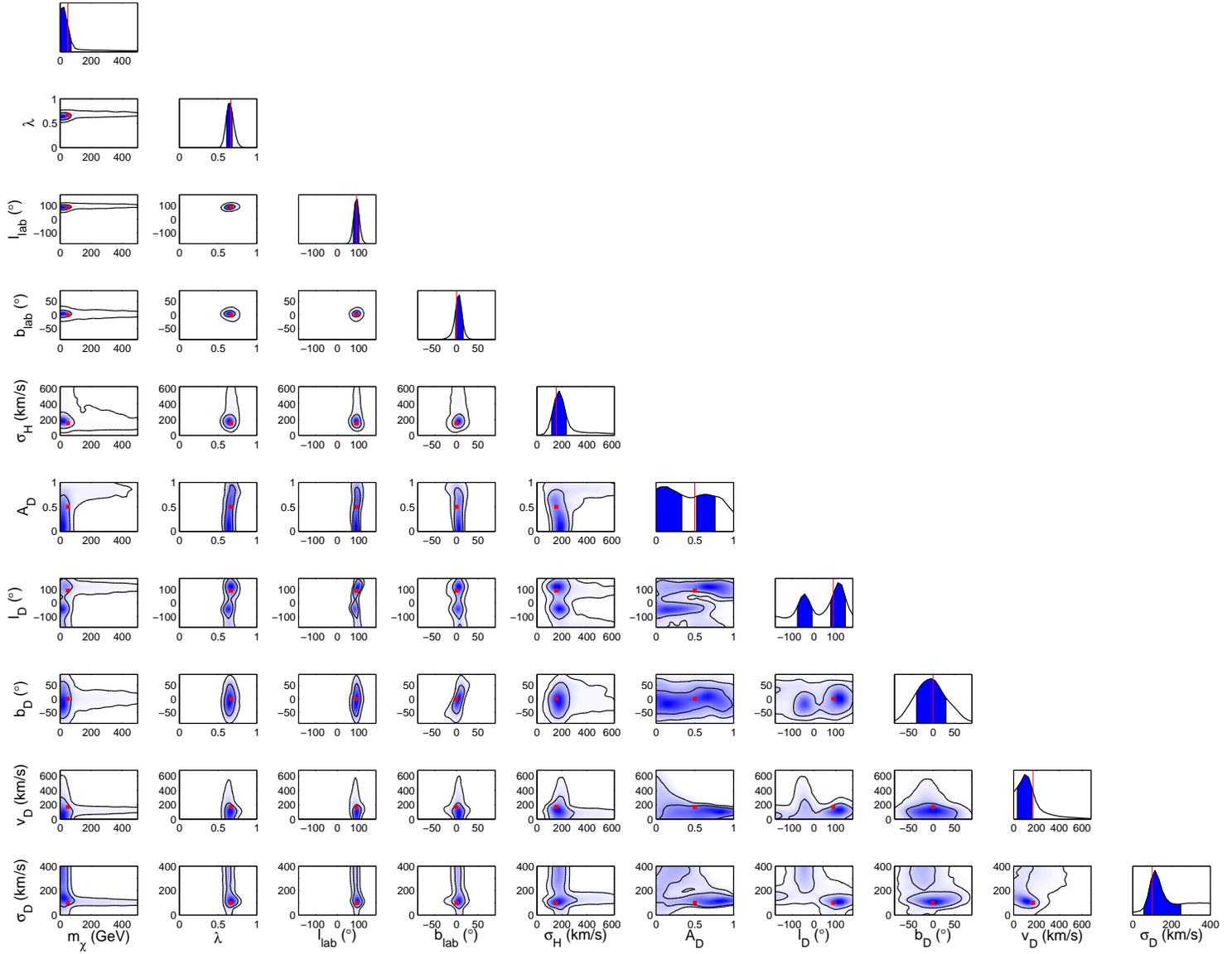}
\caption{Triangle plot for the 5--50 keV halo+disk analysis in section~\ref{subsec:halodisk}.  The disk parameters are poorly estimated, demonstrating the need for improved energy thresholds.}
\label{fig:hd-cut-tri}
\end{sidewaysfigure}

The recoil spectrum and sky maps are shown in figures~\ref{fig:hd-sp}~and~\ref{fig:hd-map}, the triangle plots for the zero-threshold and 5-keV-threshold analyses are shown in figures~\ref{fig:hd-tri}~and~\ref{fig:hd-cut-tri}, and the posterior modes and MCIs are again given in table~\ref{tab:fidparams}.  From the triangle plots, it is immediately clear that the finite energy threshold severely restricts the ability to constrain the disk-component parameters, even leading to multimodal posterior probabilities for some of the parameters.  Even in the idealized case of zero threshold the parameter estimation is imperfect, as is evidenced by the amount of bias in the estimation of the disk fraction $\AD$.  Although these results may be improved with better statistics or reduced background, it is clear that measurement of the parameters of a dark-matter disk will be challenging with current energy thresholds, if predictions of the disk parameters from N-body simulations are indeed valid.

\section{Conclusions} \label{sec:conclusions}

Using binned likelihood analyses of simulated WIMP-nucleus recoil events, we have demonstrated how detectors with directional sensitivity may place constraints on the parameters of the local dark-matter velocity distribution.  Directional sensitivity allows isotropic background events to be distinguished from signal events, but more interestingly, it also helps to break degeneracies in the velocity-distribution parameters that cannot be broken with spectral information alone.  As an illustrative example, we examined the case of a Maxwellian standard-halo-model velocity distribution.  By comparing the statistical power of energy-only, direction-only, and direction+energy analyses, we demonstrated how the degeneracy in the standard halo model between the velocity $\vlab$ of the Earth through the halo and the halo velocity dispersion $\sigmavH$ is broken with directional information.

If the local dark-matter density is indeed $\rho_0 \approx 0.3~\mathrm{GeV/cm^3}$, for a 50 GeV WIMP and a WIMP-nucleon spin-dependent cross section of $10^{-3}~\mathrm{pb}$, a MIMAC-like, 10-kg $\mathrm{CF}_4$ directional dark-matter detector might observe several hundred WIMP-nucleus recoil events in an observation period of 3 years.  With 650 signal events, the presence of a cold stream composing a fraction of the local dark-matter density may be detected, and its properties measured with relatively good accuracy.  However, the detection of a dark-disk component requires sensitivity to WIMPs moving at relatively low velocities, and may only be feasible if energy thresholds are improved. 

We have focused on an approach that seeks to maximize the amount of information about the velocity distribution, while minimizing the number of assumptions about the particle properties of the WIMP or the local density of dark matter.  The only assumptions of our approach are: 1) the form of the velocity distribution is known and can be parameterized, 2) the background is flat and isotropic.  Future work might consider the relaxation of these assumptions.  Furthermore, the analyses presented in this paper simply provide a proof of principle, showing that the parameter estimation might be feasible in cases of interest.  However, a more in-depth study of how the statistics of the parameter estimation improve as the experimental conditions are varied might be warranted, with the goal of finding the best balance between directional sensitivity and the overall event rate.  In the event that directional detectors do indeed detect WIMP-induced recoil events, such studies will be crucial in forming a complete picture of the dark-matter particle and the structure of the Galactic halo.

\begin{acknowledgments}
S.K.L. thanks Marc Kamionkowski and Jens Chluba for useful comments.  This research was supported at Caltech and Johns Hopkins University by DoE DE-FG03-92-ER40701.  A. H. G. P. is supported by a Gary McCue Fellowship through the Center for Cosmology at UC Irvine and NASA Grant No. NNX09AD09G.
\end{acknowledgments}

\begin{appendix}
\section{Event-Generation Procedure} \label{sec:eventgen}

For simple velocity distributions, eq.~(\ref{eq:spectrum}) may be used to generate random scattering events and their corresponding nuclear-recoil momentum vectors directly from the analytic form of the recoil spectrum.  However, for more general velocity distributions, an analytic form for the Radon transform may not be easily found.  In such a case, it is then easier to use the velocity distribution to generate random incoming-WIMP velocity vectors, and then to randomly generate the corresponding nuclear-recoil momentum distribution by noting that the elastic scattering is isotropic in the center-of-mass frame of the WIMP-nucleus system.

We first note that the rate for scattering by a WIMP with velocity $v$ is proportional to $v f(v)$, as can be shown from eqs.~(\ref{eq:spectrum})~and~(\ref{eq:radon}).  We thus draw randomly generated incoming-WIMP velocity vectors from the distribution proportional to $v f(\bold{v})$.  This is done using the method given in section 7.3 of ref.~\cite{press2007}.  That is, we uniformly sample points in the 4-dimensional space $(\bold{v},g)$, where the range of sampled velocity vectors $\bold{v}$ is determined by the truncation by the escape speed, and $g \in [0, \max(vf)]$ is at most the maximum value of $v f(\bold{v})$.  If a sampled point fails to satisfy the criteria $v f(\bold{v}) \geq g$, then the sampled velocity vector $\bold{v}$ is discarded; otherwise, it is retained.  The retained velocity vectors will then be distributed according to $v f(\bold{v})$, as was desired.

We then use these generated incoming-WIMP velocities to generate the corresponding nuclear recoil momenta.  We first note that the recoil direction angular distribution is isotropic in the center-of-mass frame for non-relativistic elastic scattering events; however, this clearly implies that that the lab-frame recoil direction distribution is anisotropic, and will be a function of the incoming velocity $\bold{v}$.  To be specific, it is the distribution of the angle between $\bold{v}$ and $\widehat{\bold{q}}$ that is anisotropic in the lab frame.  Thus, we could determine the appropriate distribution for each generated $\bold{v}$, and then draw the recoil directions from these anisotropic distributions.  However, it is easier to simply randomly draw the recoil direction from the isotropic center-of-mass frame distribution, and then transform it to the lab frame appropriately.

Let the direction $\widehat{\bold{v}}$ of the incoming-WIMP lab-frame velocity vector be given by the spherical coordinates $(\thetav,\phiv)$.  Furthermore, let the angle between $\widehat{\bold{v}}$ and the nuclear recoil direction $\widehat{\bold{q}}$ be $\theta$ in the lab frame, and $\thetaCM$ in the center-of-mass frame; these angles are related by
\begin{equation}
\cos \theta = \frac{1}{\sqrt{2}} (1+\cos \thetaCM)^{1/2}.
\end{equation}
Isotropic scattering in the center-of-mass frame implies that $\cos\thetaCM$ is uniformly distributed in the range $[-1,1]$; the distribution for $\cos\theta$ then follows.  However, it also implies that the angle $\phi$ between the plane defined by the lab-frame $z$-axis and $\widehat{\bold{v}}$ and the scattering plane (defined by $\widehat{\bold{q}}$ and $\widehat{\bold{v}}$) is uniformly distributed in the range $[0,2\pi]$.

Thus, we determine $\widehat{\bold{q}}$ by first randomly generating the vector $\widehat{\bold{q_0}}$ with spherical coordinates given by ${(\thetav+ \theta(\thetaCM),\phiv)}$, which lies in the $\widehat{\bold{z}}-\widehat{\bold{v}}$ plane and forms an angle $\theta$ with $\widehat{\bold{v}}$.  Rotating $\widehat{\bold{q_0}}$ by the randomly generated angle $\phi$ around the $\widehat{\bold{v}}$ axis then gives $\widehat{\bold{q}}$ with a distribution that is kinematically consistent with the incoming-WIMP velocity $\bold{v}$.  That is, $\widehat{\bold{q}} = \bold{R}(\phi,\widehat{\bold{v}})\widehat{\bold{q_0}}$, where in Cartesian coordinates
\begin{equation}
\bold{R}(\phi,\widehat{\bold{v}}) = \left( \begin{array}{lll}
\cos\phi + \widehat{v}_x^2 (1-\cos\phi) & \widehat{v}_x\widehat{v}_y(1-\cos\phi)-\widehat{v}_z\sin\phi & \widehat{v}_x\widehat{v}_z(1-\cos\phi)+\widehat{v}_y\sin\phi \\
\widehat{v}_x\widehat{v}_y(1-\cos\phi)+\widehat{v}_z\sin\phi & \cos\phi + \widehat{v}_y^2 (1-\cos\phi) & \widehat{v}_y\widehat{v}_z(1-\cos\phi)-\widehat{v}_x\sin\phi \\
\widehat{v}_x\widehat{v}_z(1-\cos\phi)-\widehat{v}_y\sin\phi & \widehat{v}_y\widehat{v}_z(1-\cos\phi)+\widehat{v}_x\sin\phi & \cos\phi + \widehat{v}_z^2 (1-\cos\phi) \end{array} \right).
\end{equation}

Finally, the nuclear-recoil momentum $q = 2\mu \cos\theta$ is given by the kinematics.  Thus, we have shown how to randomly generate the recoil momentum vectors $\bold{q} = q \widehat{\bold{q}}$ corresponding to randomly generated incoming-WIMP velocity vectors $\bold{v}$ drawn from a given velocity distribution $f(\bold{v})$.
\end{appendix}


\providecommand{\href}[2]{#2}\begingroup\raggedright\endgroup

\newgeometry{bottom=1in}
\renewcommand{\tabcolsep}{0.55mm}
\begin{sidewaystable}
\centering
\begin{tabular}{l|llllllllllll}
\hline\vspace{-.1cm}
\multirow{2}{*}{model} & $\mchi$ & \multirow{2}{*}{$\lambda$} & $\llab$  & $\blab$  & $\vlab$ & $\sigmavH$ & $\vesc$ & \multirow{2}{*}{$\AC$} & $\lC$ & $\bC$ & $\vC$ & $\sigmavC$ \\
 & (GeV)  &  & $(\circ)$ & $(\circ)$ & (km/s) & (km/s) & (km/s) &   & $(\circ)$ & $(\circ)$ & (km/s) & (km/s)\\

\hline\vspace{-.1cm}
halo-only & 50 & 1 & 90 & 0 & 220 & 155 & 550 & -- & -- & -- & -- & --\\
($\mchi$ fixed) & [50] & [1] & [90] & [0] & [0, 880] & [0, 620] & [550] & -- & -- & -- & -- & --\\

\hline\vspace{-.1cm}
halo-only & 50 & 1 & 90 & 0 & 220 & 155 & 550 & -- & -- & -- & -- & --\\
($\mchi$ flat prior) & [0, 500] & [1] & [90] & [0] & [0, 880] & [0, 620] & [550] & -- & -- & -- & -- & --\\

\hline\hline\vspace{-.1cm}
\multirow{2}{*}{halo-only} & 50 & 0.5 & 90 & 0 & 220 & 155 & 550 & -- & -- & -- & -- & --\\
 & [0, 500] & [0,1] & [0, 180] & [-90, 90] & [0, 880] & [0, 620] & [550] & -- & -- & -- & -- & --\\\cline{2-13}\vspace{-.1cm}
(6 parameters, & 62.5 & 1.00 & -81.8 & -33.7 & 168 & 207 & -- & -- & -- & -- & -- & --\\
 energy-only) & [0, 300] & [0.68, 1.0] & [-130, 110] & [-67, 25] & [35, 270] & [110, 400] & -- & -- & -- & -- & -- & --\\

\hline\hline\vspace{-.1cm}
\multirow{2}{*}{halo-only} & 50 & 0.5 & 90 & 0 & 220 & 155 & 550 & -- & -- & -- & -- & --\\
 & [0, 500] & [0, 1] & [0, 180] & [-90, 90] & [0, 880] & [0, 620] & [550] & -- & -- & -- & -- & --\\\cline{2-13}\vspace{-.1cm}
(6 parameters, & 24.4 & 0.537 & 101 & 11.0 & 215 & 136 & -- & -- & -- & -- & -- & --\\
 direction+energy) & [0, 290] & [0.45, 0.62] & [85, 110] & [2.5, 20] & [160, 280] & [98, 180] & -- & -- & -- & -- & -- & --\\

\hline\hline\vspace{-.1cm}
\multirow{2}{*}{halo+stream} & 50 & 0.684 & 90 & 0 & 220 & 155 & 550 & 0.1 & 65 & 25 & 510 & 10\\
 & [0, 500] & [0, 1] & [0, 180] & [-90, 90] & [220] & [0, 620] & [550] & [0, 1] & [0, 180] & [-90, 90] & [0, 1020] & [0, 40]\\\cline{2-13}\vspace{-.1cm}
 (direction+energy) & 78.4 & 0.647 & 88.2 & -1.76 & -- & 134 & -- & 0.118 & 67.1 & 22.9 & 460 & 13.3\\
 & [45, 200] & [0.61, 0.69] & [82, 97] & [-4.7, 4.2] & -- & [110, 150] & -- & [0.10, 0.15] & [59, 74] & [20, 27] & [430, 490] & [11, 15]\\

\hline\hline\vspace{-.1cm}
\multirow{2}{*}{halo+disk} & 50 & 0.665 & 90 & 0 & 220 & 155 & 550 & 0.5 & 90 & 0 & 170 & 100\\
 & [0, 500] & [0, 1] & [0, 180] & [-90, 90] & [220] & [0, 620] & [550] & [0, 1] & [0, 180] & [-90, 90] & [0, 680] & [0, 400]\\\cline{2-13}\vspace{-.1cm}
 (direction+energy, & 29.4 & 0.745 & 95.3 & 5.29 & -- & 134 & -- & 0.667 & 95.3 & -0.176 & 173 & 94.1\\
 zero threshold) & [3.9, 190] & [0.71, 0.78] & [85, 100] & [-2.3, 11] & -- & [110, 180] & -- & [0.46, 0.75] & [85, 110] & [-13, 12] & [140, 200] & [92, 110]\\

\hline\hline\vspace{-.1cm}
\multirow{2}{*}{halo+disk} & 50 & 0.75 & 90 & 0 & 220 & 155 & 550 & 0.5 & 90 & 0 & 170 & 100\\
 & [0, 500] & [0, 1] & [0, 180] & [-90, 90] & [220] & [0, 620] & [550] & [0, 1] & [0, 180] & [-90, 90] & [0, 680] & [0, 400]\\\cline{2-13}\vspace{-.1cm}
 (direction+energy)& 23.8 & 0.659 & 92.2 & 6.59 & -- & 177 & -- & 0.143 & 111 & -4.29 & 97.1 & 114 \\\vspace{-.1cm}
 & [0, 59] & [0.60, 0.70] & [76, 100] & [-3.8, 14] & -- & [110, 240] & -- & [0, 0.38] & [-80, -10] & [-45, 28] & [9.4, 170] & [65, 250]\\
 & & & & & & & & [0.52, 0.77] & [62, 160] & & & \\

\hline
\end{tabular}
\caption{Fiducial parameter values and flat prior ranges used to simulate data and perform likelihood analyses, respectively.  Marginalized posterior probability modes and $68\%$ minimum credible intervals are also given in a second set of rows for the halo-only 6-parameter, halo+stream, and halo+disk analyses.  Parameters for the stream and disk components (denoted by a subscript C) are both given in the 5 leftmost columns.}
\label{tab:fidparams}
\end{sidewaystable}
\restoregeometry

\clearpage

\clearpage

\clearpage

\clearpage

\clearpage

\clearpage
\clearpage

\clearpage

\clearpage

\clearpage

\clearpage

\clearpage
\end{document}